\documentclass[]{aa}
\usepackage{amsmath}
\usepackage[varg]{txfonts} 
\usepackage{graphicx}
\usepackage{subfigure}
\usepackage{color}
\usepackage{bm}
\usepackage{natbib}
\usepackage{url}
\usepackage[center, labelfont=bf, justification=justified]{caption}
\usepackage{xcolor}
\usepackage{booktabs}
\usepackage [autostyle, english = american]{csquotes}
\MakeOuterQuote{"}

\usepackage{tabularx}
\usepackage{graphicx}
\usepackage{txfonts}
\usepackage{longtable}
\usepackage{hhline}
\usepackage{arydshln}
\usepackage{multirow}
\usepackage{multicol}
\usepackage{lscape}
\usepackage{array}
\usepackage{rotating}
\usepackage{float}

\usepackage{gensymb}

\newcommand{\msun}{$M_\odot$}

\newcommand{\arcsecond}{arcsec}
\newcommand{\cmo}{cm$^{-1}$}
\newcommand{\cmt}{cm$^{-3}$}
\newcommand{\cmtw}{cm$^{-2}$}

\newcommand{\mrate}{\msun\ yr$^{-1}$}
\newcommand{\kms}{km s$^{-1}$}
\newcommand{\Ghz}{GHz}

\newcommand{\stn}{S/N}
\newcommand{\mJyBeam}{mJy beam$^{-1}$}
\newcommand{\mJBkms}{mJy beam$^{-1}$ km s$^{-1}$}
\newcommand{\klambda}{$k\lambda$}
\newcommand{\ev}{eV}
\newcommand{\kev}{keV}
\newcommand{\mev}{MeV}
\newcommand{\gev}{GeV}

\newcommand{\deutFrac}{$R_{\rm D}$}
\newcommand{\ionFrac}{$\chi_{\rm e}$}
\newcommand{\ionRate}{$\zeta_{\rm CR}$}
\newcommand{\XionRate}{$\zeta_{\rm X}$}
\newcommand{\hFrac}{$R_{\rm H}$}
\newcommand{\deplFact}{$f_{\rm D}$}

\newcommand{\bttf}{B335}
\newcommand{\bra}{19:37:00.9}
\newcommand{\bdec}{+07:34:09.6}
\newlength{\pointwidth}
\newcommand{\AnaelleProj}{2016.1.01552.S}
\newcommand{\YenProj}{2015.1.01188.S}

\newcommand{\hfs}{\textit{HfS}}

\newcommand{\co}{CO}
\newcommand{\tco}{$^{12}$CO}
\newcommand{\coseven}{C$^{17}$O}

\newcommand{\dco}{DCO$^{+}$}

\newcommand{\ntd}{N$_{2}$D$^{+}$}

\newcommand{\hco}{HCO$^{+}$}
\newcommand{\htco}{H$^{13}$CO$^+$}

\newcommand{\hydrogen}{H$_{2}$}

\newcommand{\htp}{H$_{3}^{+}$}

\begin{document} 

\title{Magnetically regulated collapse in the B335 protostar? \\II. Observational constraints on gas ionization and magnetic field coupling}
  \author{Victoria Cabedo  
               \inst{1,2,3},
              Ana\"elle Maury
               \inst{1,4},
              Josep Miquel Girart
               \inst{2,5},
               Marco Padovani
               \inst{6},
               Patrick Hennebelle
               \inst{1},
               Martin Houde
               \inst{7}
               \and
               Qizhou Zhang
               \inst{4}
               }

\institute{
Université Paris-Saclay, Université Paris Cité, CEA, CNRS, AIM, 91191, Gif-sur-Yvette, France
\and
Institut de Ci\`encies de l'Espai (ICE), CSIC,
Can Magrans s/n, E-08193 Cerdanyola del Vall\`es,  Catalonia, Spain
\and
Institute of Chemical Sciences, School of Engineering and Physical Sciences, Heriot-Watt University, EH14 4AS, Edinburgh, Scotland 
 \\
\email{v.cabedo@hw.ac.uk} 
\and
Center for Astrophysics $\mid$ Harvard \& Smithsonian, 60 Garden Street, Cambridge, MA 02138, USA 
\and
Institut d'Estudis Espacials de Catalunya (IEEC), E-08034 Barcelona, Catalonia, Spain
\and
INAF--Osservatorio Astrofisico di Arcetri, Largo E. Fermi 5, 50125 Firenze, Italy
\and
Department of Physics and Astronomy, The University of Western Ontario, 1151 Richmond Street, London, Ontario N6A 3K7, Canada
}

\abstract
{Whether or not magnetic fields play a key role in dynamically shaping the products of the star formation process is still largely debated. For example, in magnetized protostellar formation models, magnetic braking plays a major role in the regulation of the angular momentum transported from large envelope scales to the inner envelope, and is expected to be responsible for the resulting protostellar disk sizes. However, non-ideal magnetohydrodynamic effects that rule the coupling of the magnetic field to the gas also depend heavily on the local physical conditions, such as the ionization fraction of the gas.}
{The purpose of this work is to observationally characterize the level of ionization of the gas at small envelope radii and to investigate its relation to the efficiency of the coupling between the star-forming gas and the magnetic field in the Class 0 protostar \bttf.}
{We obtained molecular line emission maps of \bttf \ with ALMA, which we use to measure the deuteration fraction of the gas, \deutFrac, its ionization fraction, \ionFrac, and the cosmic-ray ionization rate, \ionRate, at envelope radii $\lesssim$1000~au. 
}
{We find large fractions of ionized gas, \ionFrac\ $\simeq 1-8 \times 10^{-6}$. Our observations also reveal an enhanced ionization that increases at small envelope radii, reaching values up to \ionRate\ $\simeq 10^{-14}$~s$^{-1}$ at a few hundred~astronomical units (au) from the central protostellar object. We show that this extreme \ionRate\ can be attributed to the presence of cosmic rays accelerated close to the protostar.}
{We report the first resolved map of \ionRate\ at scales $\lesssim 1000$~au in a solar-type Class 0 protostar, finding remarkably high values. Our observations suggest that local acceleration of cosmic rays, and not the penetration of interstellar Galactic cosmic rays, may be responsible for the gas ionization in the inner envelope, potentially down to disk-forming scales. If confirmed, our findings imply that protostellar disk properties may also be determined by local processes that set the coupling between the gas and the magnetic field, and not only by the amount of angular momentum available at large envelope scales and the magnetic field strength in protostellar cores. We stress that the gas ionization we find in B335 significantly stands out from the typical values routinely used in state-of-the-art models of protostellar formation and evolution. If the local processes of ionization uncovered in B335 are prototypical to low-mass protostars, our results call for a revision of the treatment of ionizing processes in magnetized models for star and disk formation.}


\authorrunning{V. Cabedo et al.}
\titlerunning{Deuteration and Ionization in B335}
\maketitle


\section{Introduction} 

    Observational studies have shown that magnetic fields
    are present at all scales where star formation processes are at work, permeating molecular clouds, prestellar cores, and protostellar envelopes 
    \citep[e.g.,][]{Girart2006, Alves2014, Zhang2014, Soler2019}. While magnetized models of star formation were developed early on, it is only recently that complex physics has been included in numerical magnetohydrodynamic (MHD) models, and that the predictions of dedicated models are directly compared to observations of star-forming regions. One of the major achievements of magnetized models is the prediction of the small sizes of protostellar disks, regulated by magnetic braking \citep{Dapp2012, Hennebelle2019, Zhao2020a}. Recent observations have indeed shown that disks are compact, and are almost an order of magnitude smaller than those produced in purely hydrodynamical models \citep{Maury2019, Lebreuilly2021}.
    
    In magnetized collapse models, non-ideal MHD effects play a major role in the regulation of magnetic flux during the early stages of protostellar formation \citep{Machida2011, Li2011, Marchand2020}. Hence, these processes indirectly limit the angular momentum transported by the magnetic field from large envelope scales to the inner envelope, and set the resulting disk size. However, these resistive processes depend heavily on the local physical conditions, such as dust grain properties (electric charge and size), gas density and temperature, density of ions and electrons, and the cosmic-ray (CR) ionization rate, \ionRate\ \citep[namely the number of molecular hydrogen ionizations per unit time; see e.g.,][for a review]{Zhao2020a}. Observationally, only very few measurements of these quantities have been obtained toward solar-type protostars: the effective coupling of magnetic fields to the infalling-rotating material, at scales where the circumstellar material feeds the growth in mass of the protostellar embryo and its disk, is therefore still largely unknown.

    \bttf, located at a distance of 164.5 pc \citep{Watson2020}, is an isolated Bok Globule which contains an embedded Class 0 protostar \citep{Keene1983}. The core is associated with an east--west outflow that is prominently detected in \co, with an inclination of 10\degr\ on the plane of the sky and an opening angle of 45\degr\ \citep{Hirano1988, Hirano1992, Stutz2008, Yen2010}. Its isolation and relative proximity makes \bttf\ an ideal object with which to test models of star formation. Asymmetric double-peaked line profiles observed in the molecular emission of the gas toward the envelope have been interpreted as optically thick lines tracing infalling motions, and have been extensively used to derive mass-infall rates from 10$^{-7}$ to $\sim$3$\times$10$^{-6}$ \mrate\ at radii of 100-2000~au \citep{Zhou1993,Yen2010,Evans2015}. However, new observations of optically thin emission from less abundant molecules reveal the presence of two velocity components tracing asymmetric motions at these envelope scales, which could contribute significantly to the double-peaked line profiles  \citep{Cabedo2021b}. These results suggest that simple spherically symmetric infall models might not be adequate to describe the collapse of the \bttf\ envelope, and tentatively unveil the existence of preferential accretion streamlines along outflow cavity walls.

    \bttf\ is also an excellent prototype for the study of magnetized star-formation models, because it has been proposed as an example of a protostar where the disk size is set by a magnetically regulated collapse \citep[][hereafter Paper I]{Maury2018}. This hypothesis rests mainly on two observations. First, the rotation of the gas observed at large envelope radii is not found at small envelope radii ($<$ 1000~au) and no kinematic signature of a disk was reported down to $\sim$ 10~au \citep{Kurono2013,Yen2015b}. Second, observations of polarized dust emission have revealed an "hourglass" magnetic field morphology (Paper I) in the inner envelope: comparisons to MHD models of protostellar formation \citep[see e.g.,][for a whole description of the models]{Masson2016, Hennebelle2020} suggest that the \bttf\ envelope is threaded by a rather strong magnetic field which is highly coupled to the infalling-rotating gas, preventing disk growth out to large radii. The purpose of the analysis presented here is to characterize the level of ionization and its origin in order to test the model scenario proposed in Paper I, and put observational constraints on the efficiency of the coupling of the magnetic field to the star-forming gas in the inner envelope of \bttf.
    
    In Sect.~\ref{sec:Observations}, we present the ALMA observations used to constrain physical and chemical properties of the gas at envelope radii $\lesssim$ 1000~au. In Sect.~\ref{sec:deutFrac} we derive the deuteration fraction, \deutFrac, from \dco\ (J=3-2) and \htco\ (J=3-2). In Sect.~\ref{sec:IonFrac} we compute the ionization fraction, \ionFrac, and the \ionRate. Finally, in Sect.~\ref{sec:Discussion}, we discuss our results concerning deuteration processes, the ionization, the possible influence of a local source of radiation and its effect on the coupling between the gas and the magnetic field.

\section{Observations and data reduction} \label{sec:Observations}

    Observations of the Class 0 protostellar object B335 were carried out with the ALMA interferometer during the Cycle 4 observation period, from October 2016 to September 2017, as part of the  \AnaelleProj\ project. The centroid position of \bttf\ is assumed to be $\alpha = $ \bra\ and $\delta = $ \bdec\ (in J2000 coordinates) based on the dust continuum peak obtained by \citet{Maury2018}.
    
    We used \dco\ (J=3-2) and \htco\ (J=3-2) to obtain \deutFrac\ and   \ionFrac. We used \htco\ (J=1-0) and \coseven\ (J=1-0) to derive the hydrogenation fraction, \hFrac. Additionally, we observed the dust continuum at 110 \Ghz\ to derive the estimated line opacities, CO depletion factor, \deplFact, and \ionRate. We also targeted \tco\ (J=2-1) and \ntd\ (J=3-2) for comparison purposes. All lines were targeted using a combination of ALMA configurations to recover the largest length-scale range possible. As our data for \htco\ (J=3-2) only include observations with the Atacama Compact Array (ACA), we used additional observations of this molecular line at smaller scales from the ALMA project \YenProj. Technical details of the ALMA observations are shown in Appendix~\ref{sec:ap_ObsDetails}.
    
    Calibration of the raw data was done using the standard script for Cycle 4 ALMA data and the Common Astronomy Software Applications (CASA), version 5.6.1-8. The continuum emission was self-calibrated with CASA. Line emission from \tco\ (J=2-1) and \ntd\ (J=3-2) was additionally calibrated using the self-calibrated continuum model at the appropriate frequency (231~\Ghz, not shown in this work).
    
    Final images of the data were generated from the calibrated visibilities using the tCLEAN algorithm within CASA, using Briggs weighting with a robust parameter of 1 for all the tracers. As we want to compare our data to the \coseven\ (J=1-0) emission presented in \citet{Cabedo2021b}, we adjust our imaging parameters to obtain  matching beam maps with similar angular and spectral resolution. For \dco\ (J=3-2) and \htco\ (J=3-2), we restricted the baselines to a common "u,v" range between 9 and 140~\klambda, and finally smoothed them to the same angular resolution.  The preliminary analysis shows that both \coseven\ (J=1-0) and \htco\ (J=1-0) emission is barely detected in the most extended configurations. We proceed by applying a common "uv tapering" of 1.5~\arcsec\ to both lines. This procedure allows to decrease the weighting of the largest baselines in our analysis, giving more weight to the smaller baselines and allowing us to obtain a better signal-to-noise ratio (\stn). Furthermore, we smoothed the data to obtain exactly the same angular resolution. Additionally, we smooth \coseven\ (J=1-0) and the dust continuum at 110~\Ghz\ to obtain \dco\ matching beam maps to compute \deplFact. The imaging parameters used to obtain all the spectral cubes and their final characteristics are shown in Table~\ref{table:ImageChar}. Even though the characteristics and properties of the \coseven\ (J=1-0) maps are slightly different from the ones presented in \citet{Cabedo2021b} due to the differences in the imaging process, the line profiles show the same characteristics (i.e., line profiles are double-peaked and present the same velocity pattern at different offsets from the center of the source) confirming that the imaging process has no large effect on the shape of the line profile, and that the two velocity components can still be observed.

    \begin{table*}[!ht]
        \centering
            \caption{Imaging parameters and final map characteristics.}
        \begin{tabular}{l c c c c c c c c c}
            \toprule\toprule
            & \dco\ & \htco\ & \tco\ & \ntd\ & \htco\ & \coseven** & cont. \\
                 & (J=3-2) & (J=3-2) & (J=2-1) & (J=3-2) & (J=1-0) & (J=1-0) & \\
                 \midrule
                 &&&& \\
                Rest. Freq. (\Ghz) & 216.112 & 260.255 & 230.538 & 231.321 & 86.754 & 112.359 & 110 \\
                $\Theta_{\rm LRS}$* (\arcsecond) & 11.3 & 16.0 & 10.6 & 10.6 & & & 22.3 \\ 
                Pixel size (\arcsecond) & 0.5 & 0.5 & 0.5 & 0.5 & 0.25 & 0.25 & 0.3 \\
                $\Theta_{\rm maj}$ (\arcsecond) & 1.5 & 1.5 & 1.5 & 1.5 & 2.6 & 2.6 & 0.8 \\ 
                $\Theta_{\rm min}$ (\arcsecond) & 1.5 & 1.5 & 1.5 & 1.5 & 2.6 & 2.6 & 0.7 \\ 
                P.A. (\degree) & 0 & 0 & 0 & 0 & 0 & 0 & $-61.5$ \\ 
                Spectral res. (\kms) & 0.2 & 0.2 & 0.2 & 0.2 & 0.15 & 0.15 & - \\
                rms (\mJyBeam) & 22.37 & 53.15 & 143.5 & 6.00 & 18.58 & 10.57 & 0.065  \\
                vel. range (\kms) & 7.8 - 8.9 & 7.5 - 8.9 & 7.6 - 9.4 & 7.7 - 8.9 & 7.4-9.2 & 4.7-6.5 & -   \\
                 & & & & & & 7.7-9.3 \\
                rms (\mJBkms) & 11.17 & 21.28 & 634.6 & 5.23 & 17.17 & 17.58 & -  \\
            \bottomrule
        \end{tabular}
        \begin{list}{}{}
            \item * Largest recoverable scale, computed as $\Theta_{\rm LRS} = 206265(0.6\lambda/b_{\rm min})$ in arcsec, where $\lambda$ is the rest wavelength of the line, and $b_{\rm min}$ is the minimum baseline of the configuration, both in m (\citealt{ALMAc4}).
            \item ** The two velocity ranges correspond to the two resolved hyperfine components.
        \end{list}
        \label{table:ImageChar}
    \end{table*}

    \subsubsection*{\textbf{Data products}} \label{sec:Results}
    
         From the data cubes, we obtained spectral maps that present the spectrum (in velocity units) at each pixel of a determined region around the center of the source. These maps allow us to evaluate the spectra at different offsets from the center of the envelope where the dust continuum emission peak localizes the protostar, and to detect any distinct line profile or velocity pattern. The obtained spectral maps are discussed in Appendix~\ref{sec:ap_spectralMaps}.

\begin{figure*}
            \centering
            \includegraphics[width=\textwidth]{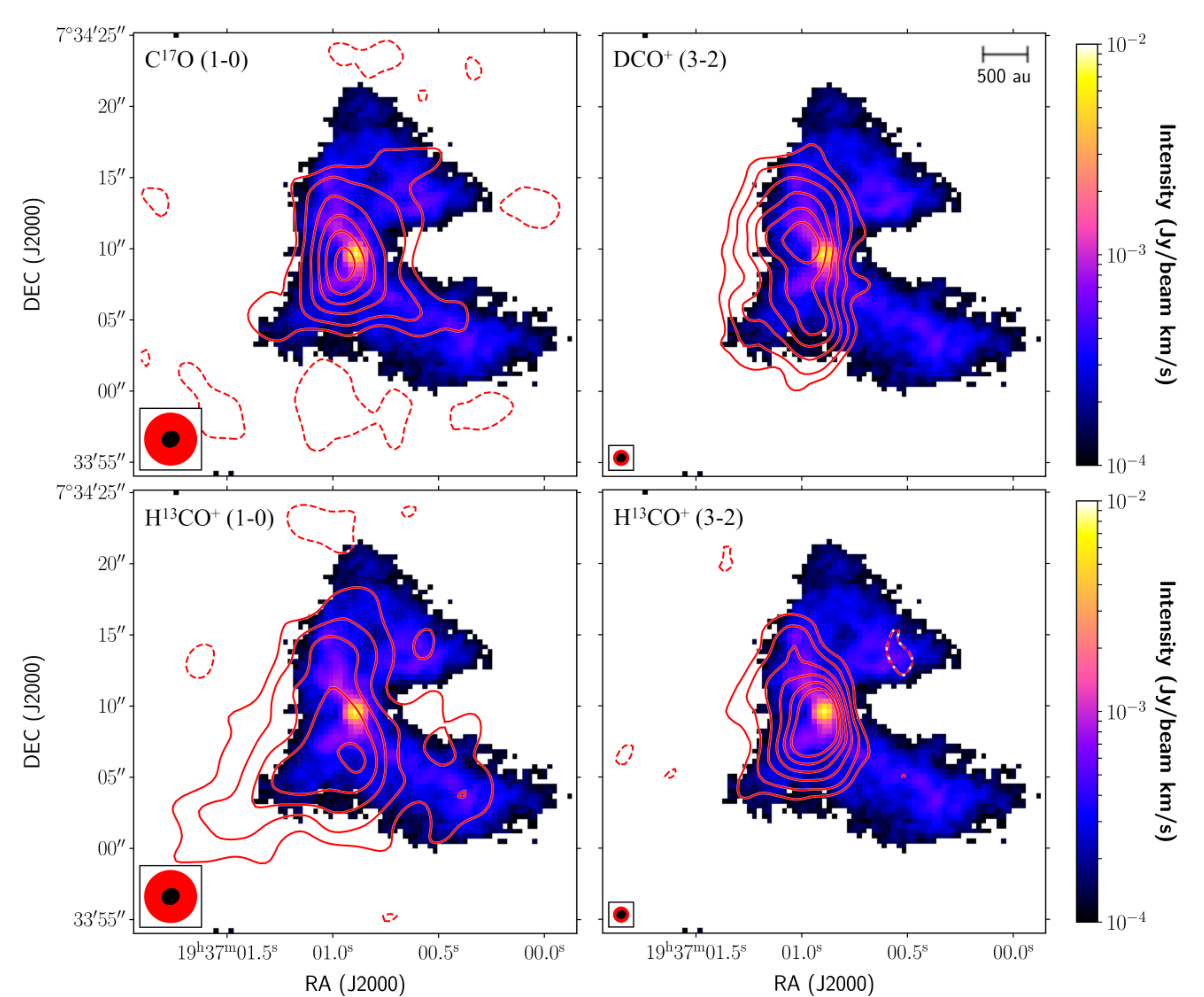}
            \caption{Map of the dust continuum emission at 110~\Ghz\ clipped for values smaller than 3$\sigma$ (see Table~\ref{table:ImageChar}) and integrated emission for the four molecular lines as indicated in the upper left corner (red contours). Contours are $-2$, 3, 5, 8, 11, 14, and 17$\sigma$ (the individual $\sigma$ values are indicated in Table~\ref{table:ImageChar}). The velocity ranges used to obtain the integrated intensity maps are shown in Table~\ref{table:ImageChar}. In the bottom-left of each plot we show the beam size for each molecular line (in red) and for the dust continuum emission (in black). The physical scale is shown in the top-right corner of the figure. 
            }
            \label{fig:intensity_maps}
        \end{figure*}

         We derived the integrated intensity maps by integrating the molecular line emission over the velocity range in which it is emitting (the velocity range used for each molecule is shown in Table~\ref{table:ImageChar}). Figure~\ref{fig:intensity_maps} shows the dust continuum emission map at 110 \Ghz\ and the integrated intensity contours of \coseven\ (J=1-0), \dco\ (J=3-2), \htco\ (J=1-0), and \htco\ (J=3-2). The \htco\ (J=1-0) emission clearly appears more spatially extended than the other lines, roughly following most of the dust emission. The emission from this line extends further than the dust toward the southeast outflow cavity wall direction. The \coseven\ (J=1-0), \dco\ (J=3-2), and \htco\ (J=3-2) emission present a similar, relatively compact morphology centered around the dust peak and elongated along the north--south equatorial plane. However, the spatial extent along the equatorial plane is slightly more compact for the \htco\ emission ($\sim$2100~au) than the \dco\ emission ($\sim$2500~au), and both are more extended than \coseven\ ($\sim$1700~au). The \dco\ peak intensity is clearly more displaced from the continuum peak than the \htco\ (J=3-2) peak; this could be a local effect around the protostar due to the large temperature and irradiation conditions, which destroy \dco\ \citep{Wootten1987, Butner1995}. The \coseven\ (J=1-0) emission appears to be slightly extended along the outflow cavities.

        As all the molecules present a double-peaked profile with similar velocity patterns to the ones observed in \citet{Cabedo2021b}, namely two different velocity components with varying intensity across the source spatial extent, we use two independent velocity components to fit the line profiles  simultaneously. Following \citet{Cabedo2021b}, we modeled the spectrum at each pixel using the \hfs\ fitting program \citep{Estalella2017}. We obtained maps of peak velocity and velocity dispersion for each velocity component, deriving the corresponding uncertainties from the $\chi^2$ goodness of fit \citep[see][for a thorough description of the derivation of uncertainties]{Estalella2017}. A more detailed discussion of these maps is presented in Appendix~\ref{sec:ap_LineMod}. In addition, an estimate of line opacities is given in Appendix~\ref{ssec:Opacity}. In the following sections, the statistical uncertainties derived (shown in Figs.~\ref{fig:deutFrac_maps}, \ref{fig:ionFrac_maps}, and \ref{fig:ionRate_maps}) are computed using standard error propagation analysis from the uncertainties of each parameter obtained with the line modeling (peak velocity, velocity dispersion, and opacity). Other systematic uncertainties are discussed separately in the corresponding sections.

\section{Deuteration fraction} \label{sec:deutFrac}
    
    The process of deuteration consists of an enrichment of the amount of deuterium with respect to hydrogen in molecular species. The deuteration fraction, \deutFrac\ = [D]/[H], in molecular ions, in particular \hco, has been extensively used as an estimator of the degree of ionization in molecular gas \citep{Caselli1998, Fontani2017}. Here, we apply this method to obtain maps of \deutFrac, computed as the column density ratio of \dco\ (J=3-2) and its non-deuterated counterpart, \htco\ (J=3-2):
    
    \begin{equation}
        R_{\rm D} = \frac{1}{f_{\rm ^{12/13}C}}  
        \frac{N_{\rm DCO^{+}}}{N_{\rm H^{13}CO^{+}}}\,,
        \label{eq:deutFrac_colDens}
    \end{equation}
    
    where $N_{i}$ is the column density of each species and $f_{^{12/13}C}$ is the abundance ratio of $^{12}$C to $^{13}$C. The column density of both species is computed such as:
    
    \begin{equation}
        \begin{split}
        N_i = &\frac{8 \pi}{\lambda^{3}A} \frac{1}{J_{\nu}(T_{\rm ex}) - J_{\nu}(T_{\rm bg})}\times \\ & \frac{1}{1-\exp(-h\nu/k_{\rm B}T_{\rm ex})} \frac{Q_{\rm rot}}{g_{u}\exp(-E_{l}/k_{\rm B}T_{\rm ex})}\times \\ & \int{I_{\rm 0} {\rm d} v},
        \end{split}
        \label{eq:colDens}
    \end{equation}

    where $\lambda$ is the wavelength of the transition, $A$ is the Einstein coefficient, $J_{v}(T)$ is the Planck function at the background (2.7~K) and at the excitation temperature of the line (assumed to be equal to the dust temperature at the observed scales, $T_{\rm ex}=T_{\rm d}=25$~K, \citealt{Galametz2019}, \citealt{Cabedo2021b}), $k_{\rm B}$ is the Boltzmann constant, $g_{u}$ is the upper state degeneracy, $Q_{\rm rot}$ is the partition function at 25~K, $E_{l}$ is the energy of the lower level, and $\int{I_{0} {\rm d} v }$ is the integrated intensity. Values of these parameters for each molecule are listed in Table~\ref{tab:colDens_pars}.
    
    \begin{table}
        \centering
        \caption{Parameters for the computation of the column density.}
        \begin{tabular}{l c c c}
            \toprule\toprule
            Transition & \dco\ (J=3-2) & \htco\ (J=3-2) & \coseven\ (J=1-0) \\
            \midrule
            $\log~(A/{\rm s}^{-1})$ & $-3.12$ & $-2.87$ & $-7.17$ \\
            $g_{u}$ & 7 & 7 & 3\\
            $Q_{\rm rot}$ & 25.22 & 22.91 & 25 \\
            $E_{l}$ (\cmo) & 7.21 & 8.68 & 0.00 \\
            $N_{i}^{\rm mean}$ (\cmtw) & 3$\times$10$^{11}$ & 7$\times$10$^{11}$ & 3$\times$10$^{13}$ \\
            \bottomrule
        \end{tabular}
        \label{tab:colDens_pars}
    \end{table}
    
\begin{figure*}
        \centering
        \includegraphics[width=\textwidth]{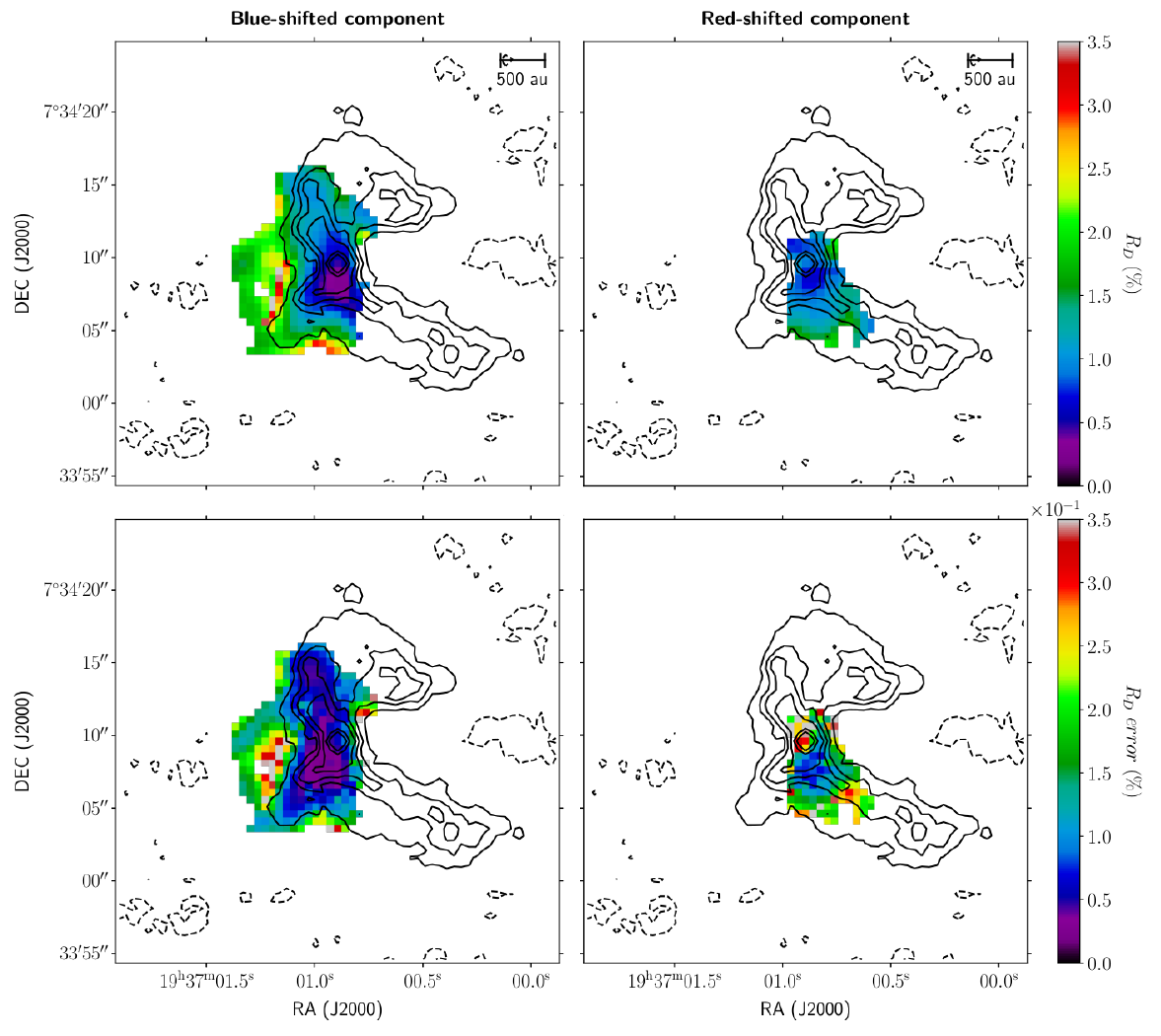}
        \caption{Dust continuum emission at 110 \Ghz\ for $-2$, 3, 5, 7, 10, 30, and 50$\sigma$ (black contours) superimposed on the deuteration fraction maps (top panels) and on the corresponding statistical uncertainties (lower panels) for the blue- and redshifted velocity components (left and right column, respectively). The spatial scale is shown in the top-right corner of the upper panels.}
        \label{fig:deutFrac_maps}
    \end{figure*}

    The top panels of Fig.~\ref{fig:deutFrac_maps} show the \deutFrac\ maps of \bttf\ for the blue- and redshifted velocity components (left and right column, respectively). The mean values of \deutFrac\ are $\sim$1-2\%, being higher in the outer region of the envelope where the gas is expected to be colder, and decreasing toward the center where the protostar is located and the temperature is expected to rise. The bottom panels show that the statistical uncertainties are roughly one-tenth of the deuteration values. We note that the highest values of \deutFrac\ are found for the blueshifted component, as high as 3.5\%. However, these values have the largest associated errors due to the presence of a third velocity component in the line profiles of \htco\ (J=3-2) and should not be totally trusted (see Appendix~\ref{sec:ap_spectralMaps}). Finally, we note that while the average \deutFrac\ values for both velocity components are similar, the most widespread blueshifted gas component shows larger \deutFrac\ dispersion, with values from 0.06\%\ to 2\%, while the more localized redshifted gas exhibits more uniform values, between 0.01\% and 1\%. 
    
    One of the main uncertainties of the deuteration analysis is the assumption that both \dco\ (J=3-2) and \htco\ (J=3-2) have the same $T_{\rm ex}$. This should be accurate enough, because both molecules present a similar geometry and dipolar moment and are assumed to be spatially coexistent. However, to check the effect that the variation of $T_{\rm ex}$ would produce in our values, we computed the column densities of both species at two additional temperatures, 20 and 30~K. While the \dco\ column density does not change significantly at these two temperatures with respect to the value adopted (25~K), the \htco\ column density changes by about 15\%. Nevertheless, the new values are still in agreement with the derived values of \deutFrac\ from \dco\ reported in the literature (see Sect.~\ref{sec:deuteration_discussion}). 
        
\section{Ionization fraction and cosmic-ray ionization rate} \label{sec:IonFrac}

    We followed the method presented in \citet{Caselli1998} to compute \ionFrac\ and \ionRate\ from \deutFrac. The two quantities are given by
    
    \begin{equation}
        \chi_{\rm e} = \frac{2.7\times10^{-8}}{R_{\rm D}} - \frac{1.2\times10^{-6}}{f_{\rm D}},
        \label{eq:ionization_frac}
    \end{equation}
    
    and
    
    \begin{equation}
        \zeta_{\rm CR} = \left[ 7.5\times10^{-4} \chi_{{\rm e}} + \frac{4.6\times10^{-10}}{f_{{\rm D}}} \right] \chi_{{\rm e}} n_{{\rm H_{2}}} R_{{\rm H}}\,,
        \label{eq:ionization_rate}
    \end{equation}
    
    where \deplFact\ is the depletion fraction of C and O, $n_{\rm H_{2}}$ is the \hydrogen\ volume density, and \hFrac\ is the hydrogenation fraction, \hFrac\ = [\hco]/[\co].
    
    While this observational method to measure \ionFrac\ of the gas relies on a simple chemical network and a rather strong hypothesis, it is widely used in the literature in dense cores and also toward protostars. The validity and shortcomings of the chemical models used to constrain ionization fractions are concisely discussed in Sect.~\ref{sec:ionization_discussion}.

    \subsection{Estimate of the CO depletion in B335} \label{sec:depletion}
    
        As \deutFrac, and therefore \ionFrac, depend on the level of C and O depletion from the gas phase, we hypothesize that the CO abundance is proportional to the CO column density and estimated \deplFact\ as the ratio between the "expected", $N_{\rm CO}^{\rm exp}$, and the "observed" \co\ column density, $N_{\rm CO}^{\rm obs}$:
        
        \begin{equation}
            f_{\rm D} = \frac{N_{\rm CO}^{\rm exp}}{N_{\rm CO}^{\rm obs}}\,.
            \label{eq:depletionFrac}
        \end{equation}

        Here, $N_{\rm CO}^{\rm exp}$ is computed as the product of the \hydrogen\ column density, $N_{\rm H_2}$, and the expected \co\ to \hydrogen\ abundance ratio, $X_{\rm CO}$ = [\co]/[\hydrogen] = 10$^{-4}$ \citep{WilsonRodd1994,Gerner2014}, and $N_{\rm CO}^{\rm obs}$ is derived as the product of the \coseven\ column density, $N_{\rm C^{17}O}$, and the \coseven\ to \co\ abundance ratio, $f_{\rm C^{17}O}$ = [\co]/[\coseven] = 2317 \citep{Wouterloot2008}. Then, \deplFact\ is given by

        \begin{equation}
            f_{\rm D} = \frac{N_{\rm H_2} X_{\rm CO}}{N_{\rm C^{17}O} f_{\rm C^{17}O}}\,.
            \label{eq:depletion}
        \end{equation}

        The \hydrogen\ column density has been estimated from the dust thermal emission at 110 \Ghz\ (see Fig.~\ref{fig:intensity_maps}) corrected for the primary beam attenuation. Indeed, $N_{\rm H_2}$ depends directly on the flux density measured on the continuum map, the beam solid angle $\Omega_{\rm beam}$, the Planck function $B_{\nu}(T_{\rm d})$ at the dust temperature, and the dust mass opacity $\kappa_{\nu}$ (in units of cm$^2$ g$^{-1}$) at the frequency at which the dust thermal emission was observed: 

       \begin{equation}
           \label{eq:column-density}
           N_{\rm H_2} = -\frac{1}{\mu_{\rm H_2}  m_{\rm H}  \kappa_{\nu}}  \ln \left[ 1 - \frac{S_{\nu}^{\rm beam}}{\Omega_{\rm beam} B_{\nu}(T_d)}  \right]\,,
       \end{equation}
        where $\mu_{\rm H_2}=2.8$, $m_{\rm H}$ is the atomic hydrogen mass, and for the dust opacity we used a power-law fit given by

        \begin{equation}
            \label{eq:kappa}
            \kappa_{\nu} = \frac{\kappa_0}{\chi_{\rm d}} \left( \frac{\nu}{\nu_0} \right) ^{\beta}.
        \end{equation}
        We adopt a dust mass absorption coefficient $\kappa_0 = 1.6$~cm$^2$~g$^{-1}$ at $\lambda_0 = 1.3$~mm, following \citet{Ossenkopf1994}. We assume a standard gas-to-dust ratio $\chi_{\rm d}$ = 100, a dust emissivity exponent $\beta=0.76$ \citep{Galametz2019}, and a dust temperature $T_{\rm d}=25$~K. The resulting $N_{\rm H_2}$ column densities probed in the ALMA dust continuum emission map range from $\sim 10^{20}$ cm$^{-2}$ at radii of $\sim 1600$~au up to a few $10^{22}$ cm$^{-2}$ at the protostar position. We note that the values of column density suffer from a systematic error due to assumptions on the parameters used to derive them (e.g., dust opacity, dust temperature). We estimate that the systematic error is about a factor of 3 to 5, being higher toward the peak position where standard dust properties and conditions of emission may not be met. We use Eq.~\ref{eq:colDens} to estimate the observed CO column density $N_{\rm CO} = N_{\rm C^{17}O} f_{\rm C^{17}O}$ from the C$^{17}$O emission map shown in Fig.~\ref{fig:intensity_maps}, assuming an excitation temperature  of 25~K. The CO column density values are subject to systematic uncertainties due to the adopted gas temperature, as this latter is not well constrained and is expected to vary across the envelope. Altogether, the systematic uncertainty is about a factor of 2, and is higher toward the peak position where different gas temperatures are expected to be present along the line of sight.
        
\begin{figure}
            \centering
            \includegraphics[width=9cm]{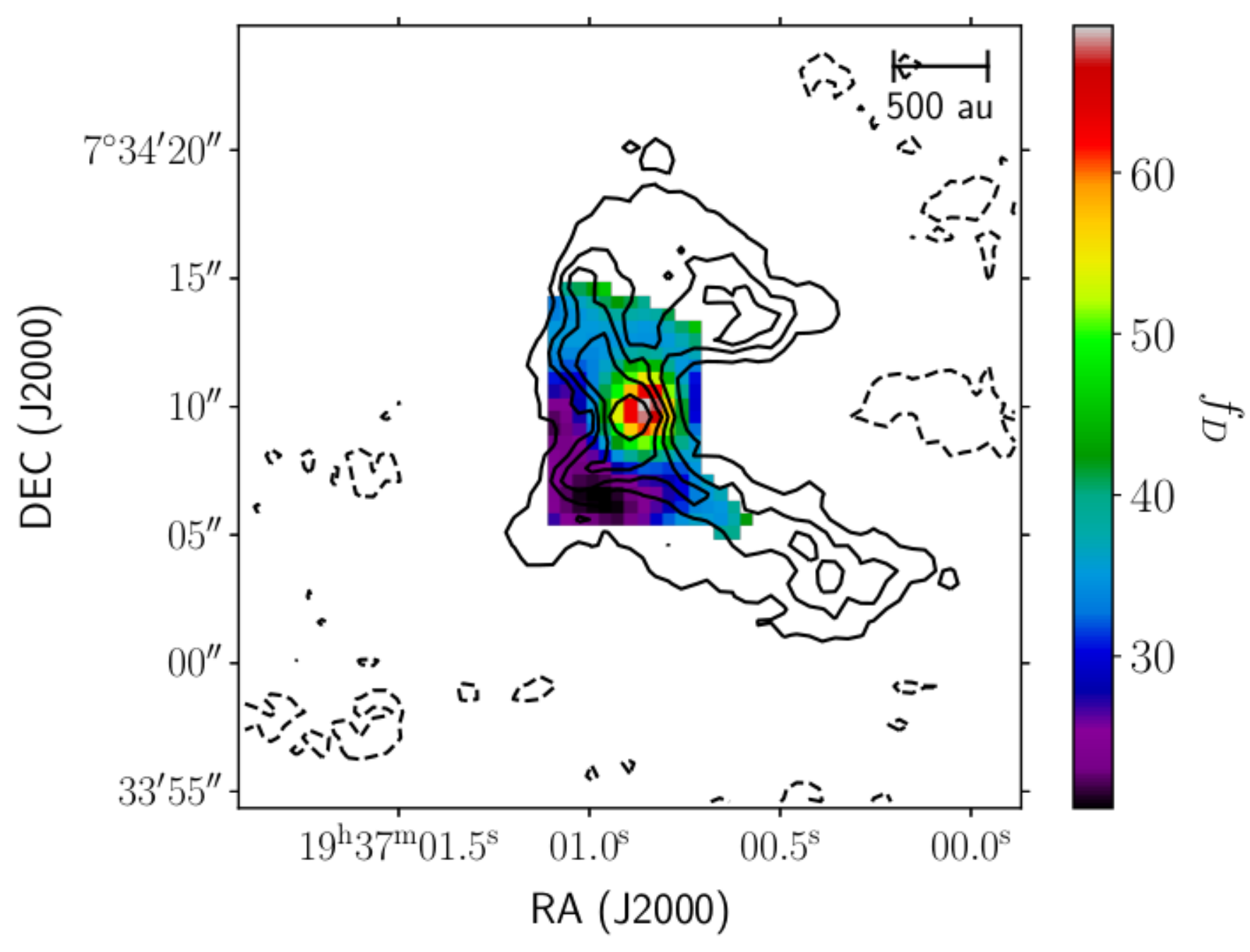}
            \caption{Map of the CO depletion factor and contours of dust continuum emission at 110~\Ghz, for emission over 3$\sigma$. The spatial scale is shown in the top-right corner of the figure.}
            \label{fig:depletionFrac}
        \end{figure}

        Using Eq.~\ref{eq:depletion}, we obtain the CO depletion fraction map shown in Fig.~\ref{fig:depletionFrac}. The depletion factor ranges from $\sim 20$ in the outer regions to $\sim 70$ in the center of the object. Propagating the errors from the column density maps they stem from, we expect these depletion values to have an uncertainty of a factor of 2 to 3, as the main source of errors ---which are due to the assumed gas and dust temperatures--- are lessened by the ratio. The CO depletion seems highly asymmetric around the protostar, being the lowest in the southeastern quadrant, while high values are associated to the protostar position and the northern region. We note that the regions with high \deplFact\ coincide with the regions of low deuteration shown in Fig.~\ref{fig:deutFrac_maps}. 
        
    \subsection{Ionization fraction of the gas} \label{sec:IonFrac_derivation}

\begin{figure*}
            \centering
            \includegraphics[width=\textwidth]{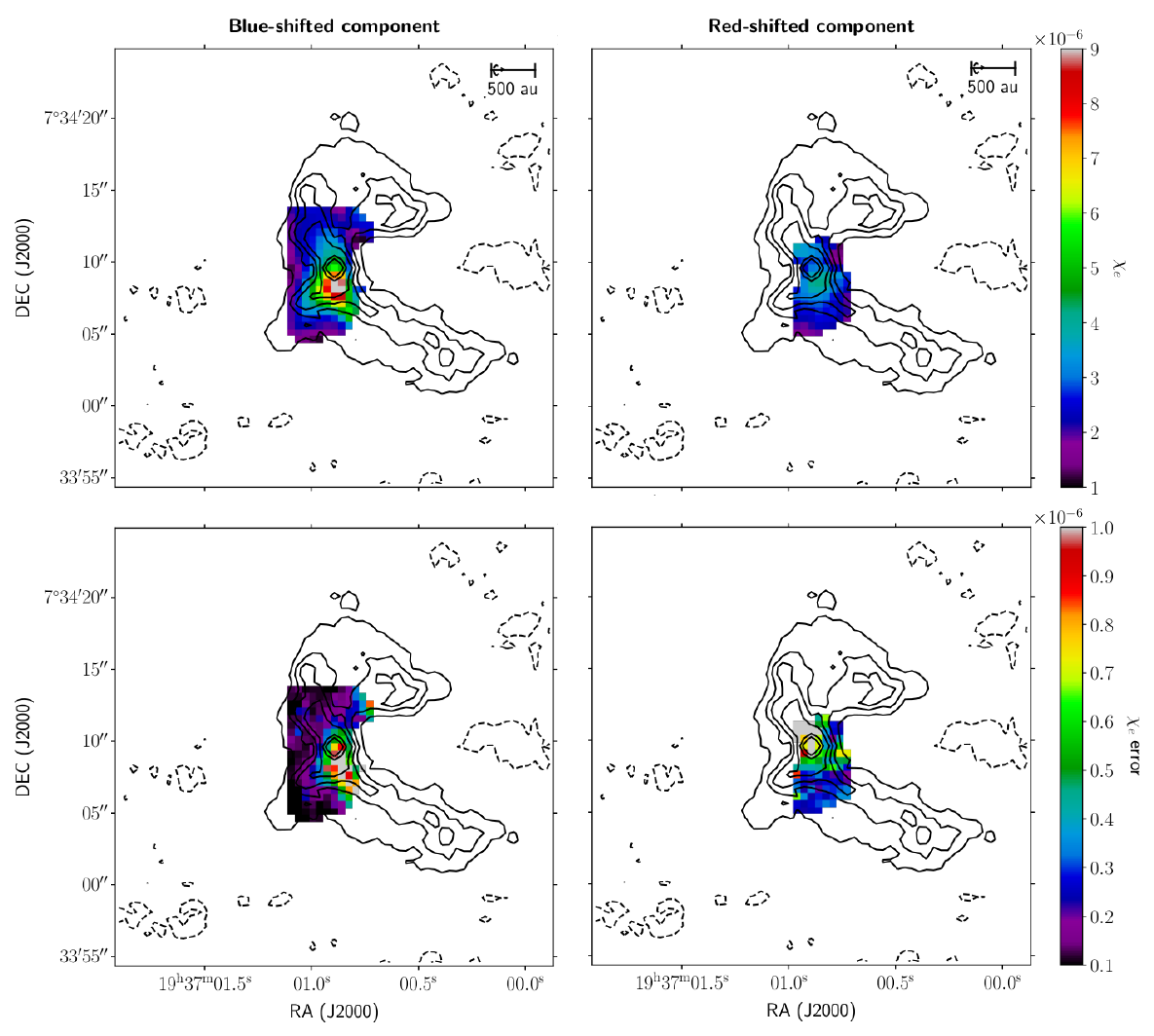}
            \caption{Dust continuum emission at 110~\Ghz\ for $-2$, 3, 5, 7, 10, 30, and 50$\sigma$ (black contours) superimposed on the ionization fraction maps (top panels) and on the corresponding statistical uncertainties (lower panels) for the blue- and redshifted velocity components (left and right column, respectively). The spatial scale is shown in the top-right corner of the upper panels.}
            \label{fig:ionFrac_maps}
        \end{figure*}

        We obtained \ionFrac\ using our \deutFrac\ maps, shown in Sect.~\ref{sec:deutFrac} and the CO depletion map, presented in Fig.~\ref{fig:depletionFrac}. The top panels of Fig.~\ref{fig:ionFrac_maps} show the derived \ionFrac\ and the bottom panels the associated uncertainties, for the blue- and red-shifted velocity components (left and right column, respectively).  We obtained a mean value of \ionFrac\ = $2\times10^{-6}$ for both components. Generally, these values are subject to statistical uncertainties of $\sim1.5\times10^{-6}$. The range of values found for \deutFrac\ and \deplFact\ indicates that the \ionFrac\ depends mostly on the level of deuteration, and therefore the systematic errors on the estimated \ionFrac\ from Eq.~\ref{eq:ionization_frac} are typically $<30\%$ and statistical uncertainties dominate in this case. As \deutFrac\ decreases toward the center of the source, we find that \ionFrac\ increases toward the center. 
        
    \subsection{Derivation of the CR ionization rate} \label{sec:IonRate}
    
        We produced \ionRate\ maps using Eq.~\ref{eq:ionization_rate}. We obtained $n_{\rm H_{2}}$ as $N_{\rm H_{2}}/L$, using the $N_{\rm H_{2}}$ map obtained in Sect.~\ref{sec:depletion} and considering a core diameter, $L$, of 1720~au derived from the \coseven\ (J=1-0) emission \citep{Cabedo2021b}. With this approximation, we find values of the volume density ranging from $\sim 10^{4}$ in the outer regions, to $\sim 10^{6}$ \cmt\ at the protostar position. We obtained \hFrac\ for both velocity components as the column density ratio of \htco\ (J=1-0) to \coseven\ (J=1-0), and accounting for $f_{^{12/13}C}$ and $f_{\rm C^{17}O}$:
        
        \begin{equation}
            R_{\rm H} = \frac{N_{\rm H^{13}CO^{+}} f_{^{12/13}C}}{N_{\rm C^{17}O} f_{\rm C^{17}O}}\,.
        \end{equation}
        Values of \hFrac\ are relatively uniform across the source, with mean values between 2 and $3\times 10^{-7}$ for the two velocity components. The statistical uncertainties of these values are generally one order of magnitude smaller than \hFrac.
        
\begin{figure*}
            \centering
            \includegraphics[width=\textwidth]{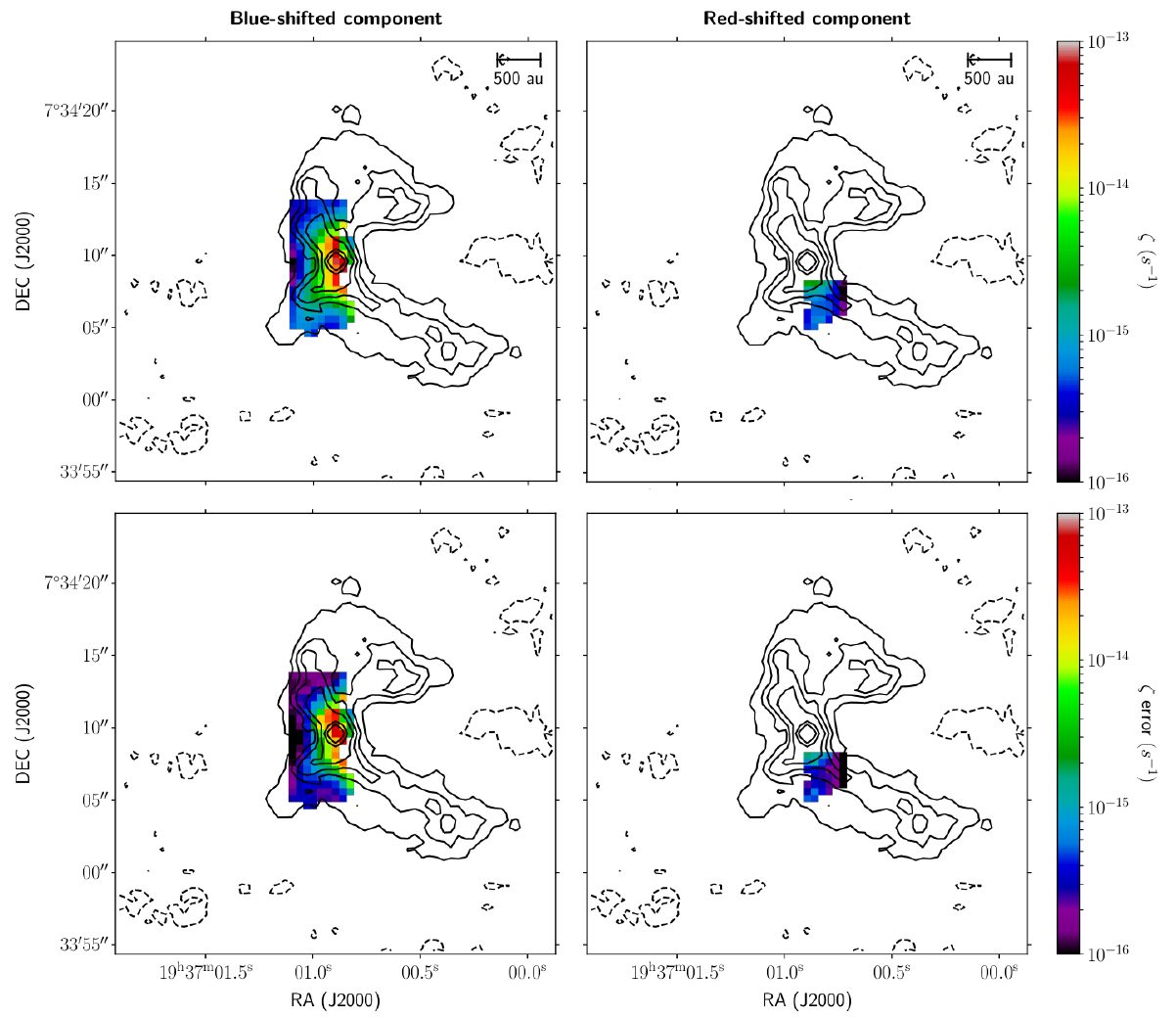}
            \caption{Dust continuum emission at 110~\Ghz\ for $-2$, 3, 5, 7, 10, 30, and 50$\sigma$ (black contours) superimposed on the ionization rate maps (top panels) and on the corresponding statistical uncertainties (lower panels) for the blue- and redshifted velocity components (left and right column, respectively). The scale is shown in the top-right corner of the upper panels.}
            \label{fig:ionRate_maps}
        \end{figure*}

        The two terms within the brackets in Eq.~\ref{eq:ionization_rate} have values of the order of $\sim10^{-9}$ and $\sim10^{-11}$, respectively. Therefore, the derivation of \ionRate\ is very sensitive to \ionFrac\ but not to \deplFact. The top panels of Figure~ \ref{fig:ionRate_maps} show the derived \ionRate\ and the bottom panels show the associated statistical uncertainties, for the blue- and redshifted velocity components (left and right column, respectively). As expected from the values of \ionFrac, \ionRate\ increases toward the center, reaching values of 7$\times$10$^{-14}$~s$^{-1}$ at the peak of the dust continuum. The uncertainties on \ionRate\ are large, due to all the errors coming from the modeling of the different molecules. Considering the systematic uncertainties on n$_{\rm H_{2}}$, and \hFrac\, we estimate the global uncertainty on \ionRate\ values, and find them to be of the same order of magnitude as the \ionRate\ values. 
        
\begin{figure*}
            \centering
            \resizebox{.85\hsize}{!}{
            \includegraphics[]{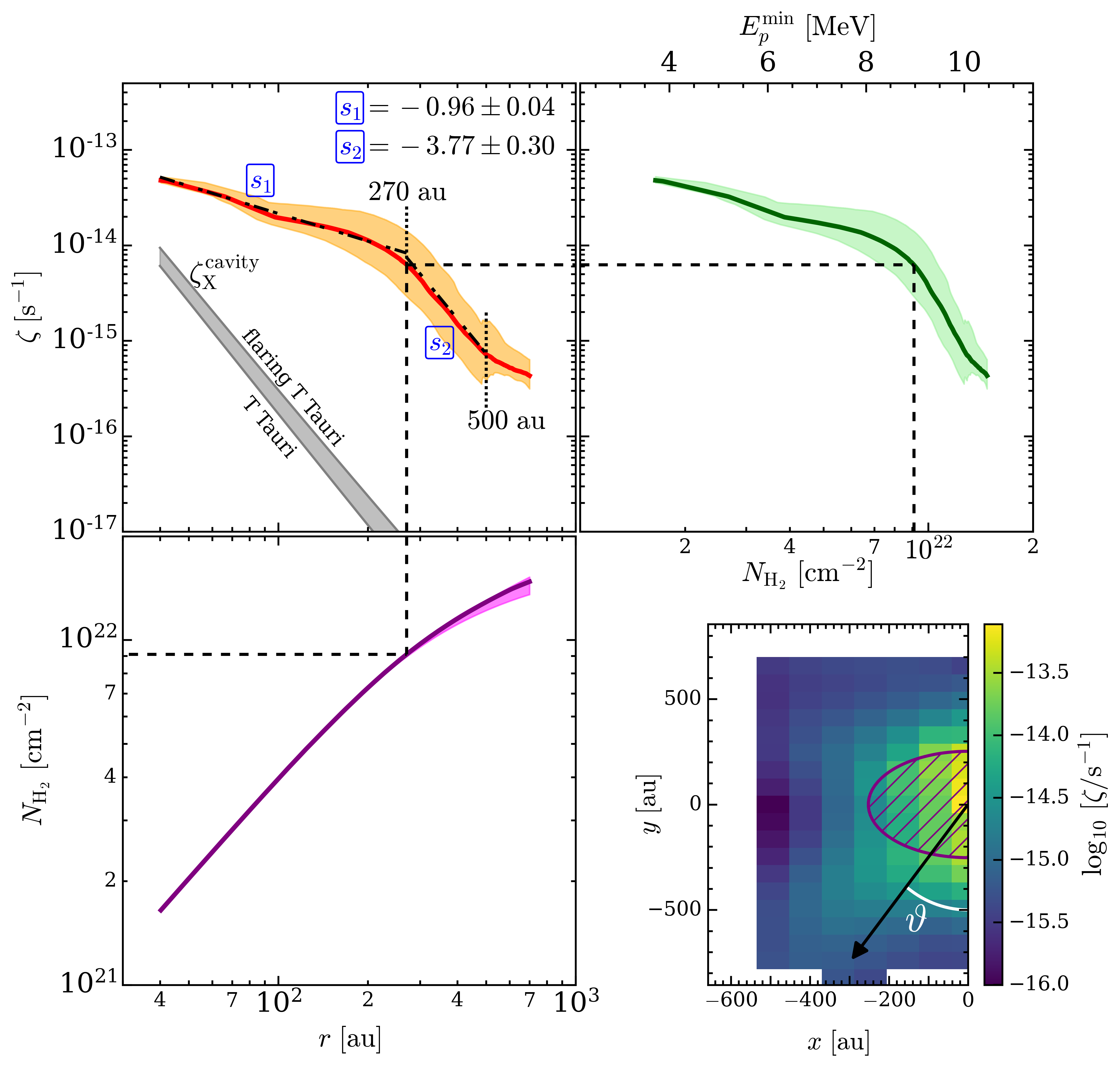}}
            \caption{{\em Upper left panel}: \ionRate\ as a function of the radius, $r$. The two dotted black lines at 270 and 500~au identify the radius ranges used to compute the slopes ($s_{1}$ and $s_{2}$, respectively). The gray shaded region shows the expected X-ray ionization in the outflow cavity for a typical and a flaring T Tauri star \citep{Rab2017}. {\em Lower left panel}: Hydrogen column density, ${\bm N_{\rm H_{2}}}$, from the source center outwards as a function of radius. {\em Upper right panel}: \ionRate\ as a function of the \hydrogen\ column density. The upper x-axis shows the minimum energy that CR protons must have in order not to be thermalized \citep{Padovani2018}. The orange, magenta, and green filled regions in the three upper panels represent the envelope of the \ionRate\ and ${\bm N_{\rm H_{2}}}$ profiles (see Sect.~5.2), while the red, purple, and dark green solid lines show their median value. {\em Lower right panel}: \ionRate\ map of the blueshifted component (zoom of the upper left panel of Fig.~5) in logarithmic scale (colored map). The purple hatched region shows a circle of radius 270~au, where CRs propagate according to the diffusive regime. For illustration purposes, the black arrow at the position angle $\vartheta$ shows a direction used to extract the \ionRate\ profile from the map. The black dashed lines identify the radius as well as the corresponding values of ${\bm N_{\rm H_{2}}}$ and \ionRate, where the slope of \ionRate\ changes from $s_{1}$ to $s_{2}$.
            }
            \label{fig:zeta_vs_radii}
        \end{figure*}

    To interpret the trend of \ionRate\ as a function of the distance from the density peak in an unbiased way, we considered profiles of \ionRate\ along directions from the density peak outward denoted by the position angle $\vartheta$, with $0\le\vartheta\le\pi$ (see the lower right panel of Fig.~\ref{fig:zeta_vs_radii}). The envelope of the profiles is represented by the orange filled region in the upper left panel of Fig.~\ref{fig:zeta_vs_radii}. As the \ionRate$(r,\vartheta)$ distributions between 40 and 700~au are skewed, for each radius $\bar{r}$ we considered the median value of \ionRate$(\bar{r},\vartheta)$ and estimated the errors using the first and third quartiles. We find that the trend of \ionRate\ can be parameterized by two independent, decreasing power-law profiles \ionRate$(r)\propto r^{s}$. At small radii, $r<270$~au, the slope of the power law, $s = -0.96 \pm 0.04$, is compatible with the diffusive regime of CR propagation. However, at $r>270$~au, \ionRate\ drops abruptly to $s = -3.77 \pm 0.30$, which cannot be explained by any propagation regime. The possible explanations for this result are explained with more details in Sect.~\ref{sec:ionization_discussion}.
        
\section{Discussion} \label{sec:Discussion}

    \subsection{Local destruction of deuterated molecules} \label{sec:deuteration_discussion}
    
        The deuteration fractions observed in Class 0 protostars range between 1\%\ and 10\%, with a large scatter from source to source and at the different protostellar scales \citep{Caselli2002,Roberts2002,Jorgensen2004}. The [\dco]/[\htco] values that we find for \bttf\  are therefore in broad agreement with the typical range reported in the literature, even accounting for temperature effects. However, \citet{Butner1995} found a [\dco]/[\hco] of $\simeq$3\% in B335 at scales of $\sim$10000~au. This value is larger than the range of values we find with ALMA at smaller scales, which indicates a decrease in \deutFrac\ toward the center of the envelope. Indeed, \deutFrac\ decreases down to $<1$\% toward the center and the northern region of the source. This decrease could be attributed to local radiation processes occurring during the protostellar phase, after the protostar has been formed. An increase in the local radiation could promote processes that would lead to the destruction of deuterated molecules due to the evaporation of neutrals to the gas phase \citep{Caselli1998, Jorgensen2004}, or the increased abundance of ionized molecules like \hco\ \citep{Gaches2019}.
   
\begin{figure*}
            \centering
            \includegraphics[width=0.99\textwidth]{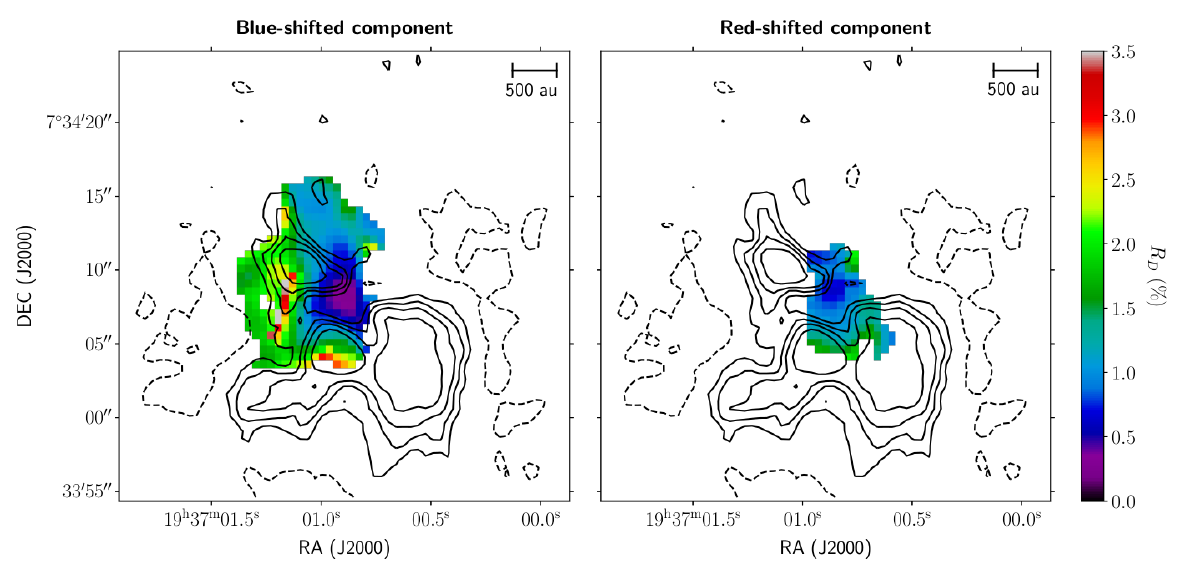}
            \caption{Deuteration fraction maps for the blue- and redshifted velocity components (left and right column, respectively). The \ntd\ (J=3-2) integrated emission is shown overlaid as black contours at 3, 5, 10, 15, and 20$\sigma$, where $\sigma$ = 5.23 \mJyBeam. The spatial scale is shown in the top-right corners of each panel.}
            \label{fig:deutFrac_n2d_maps}
        \end{figure*}

        The decreasing trend in \deutFrac\ is further confirmed by our observations of \ntd\ (J=3-2). Figure \ref{fig:deutFrac_n2d_maps} shows the deuteration map overplotted with the \ntd\ (J=3-2) integrated emission. A lack of \ntd\ is observed in regions where \deutFrac\ is lower, suggesting that the physical processes lowering the abundance of O-bearing deuterated molecules also lower the abundance of N-bearing deuterated species around the protostar.

    \subsection{High values of C and O depletion  in the source} \label{sec:depletion_discussion}
            
        At core scales, observations of the depletion in \bttf\ have been rather inconclusive as some studies suggest very little CO depletion (JCMT 20-arcsec beam probing 3000~au, \citealt{Murphy1997}), while others suggest a CO depletion of one order of magnitude, \deplFact $\sim 10$, at similarly large scales \citep{Walmsley1987}.  Our spatially resolved observations reveal large values of CO depletion  of the gas in the inner envelope of \bttf\  at radii $<1000$~au. As shown in Fig.~\ref{fig:depletionFrac}, we find \deplFact\ ranging from 20 to 40 at radii of 1000 to 100~au, except at the core center where it increases up to 80. Such CO depletion at protostellar radii $100-1000$~au is somehow unexpected, if one assumes depletion is due to freeze-out onto dust grains. Indeed, gas and dust at such radii should have temperatures beyond $\sim20-30$~K, 
        where the evaporation of CO from the surface of the dust should replenish it in the gas phase \citep{Anderl2016}. The high depletion values found at the protostar position are highly uncertain due to unknown optical depth, and unconstrained dust and gas temperatures. We stress that the possible depletion increase toward the center in B335 should be confirmed with further observations of higher J transitions of CO.

        Several observational studies of protostellar cores have found depletion factors $f_D>10$ at scales of $\sim 5000$~au, with some spatially resolved studies pointing to values up to  $\sim 20$ \citep{Alonso-Albi2010,Christie2012,Yildiz2012,Fuente2012}. Models of protostellar evolution with short warm-up timescales in the envelope \citep[see e.g.,][]{Aikawa2012} show that \deplFact\ may remain high even inside the sublimation radius, especially at early stages, because of the conversion of CO into CH$_{3}$OH and CH$_{4}$ on the grain surfaces. Photodissociation due to the protostellar UV radiation could contribute to lowering the global CO abundance in the innermost layers of protostellar envelopes \citep{Visser2009b}. Additionally, other sources of ionization, such as CRs, could decrease the abundance of \co\ in the gas phase by promoting its reaction for the formation of \hco\ \citep{Gaches2019}. This decrease in CO by transformation into \hco\ could also explain the observed coincidence between regions with large CO depletion and low \deutFrac.

        Another possible cause for the observed depletion could be the presence of relatively large dust grains, as observed in several protostars at similar scales by \citet{Galametz2019}. Indeed, larger grains are less efficiently warmed up \citep{Hiqbal2018}, limiting the desorption of CO ices to the gas phase. We note that the presence of large dust grains in the inner envelope of \bttf\ could also favor enhanced ionization of the gas \citep{Walmsley2004}, and thus also be consistent with our findings regarding ionization.

    \subsection{Origin of the ionization}
    \label{sec:ionization_discussion}
    
        From the single-dish observations of \citet{Butner1995}, \citet{Caselli1998} find ionization fractions of protostellar gas
        at core scales of the order of 10$^{-8}$--10$^{-6}$. In the B335 outer envelope where $n(\rm {H_2})  \sim 10^{4}$ cm$^{-3}$, these authors estimate \ionFrac\ of a few $10^{-6}$. With our interferometric data, we measure \ionFrac\ $\sim$ 2$\times$10$^{-6}$ in the inner envelope, where $n(\rm {H_2}) \sim 10^6$ cm$^{-3}$,  with the largest values being \ionFrac\ $\sim$9$\times$10$^{-6}$. We note that while all these aforementioned values are only providing line-of-sight averages for \ionFrac, the interferometric data only select a narrow range of spatial scales, making them less sensitive to line-of-sight averaging than single-dish data, and therefore they are more representative of local abundance variations.
        
        We stress that, while the use of deuteration to measure \ionFrac\ is a standard practice in the literature, it has been noted that such an approach may not give accurate results in low-density medium experiencing small deuteration fractions (e.g., $\sim 10^{3} \rm{cm}^{-3}$, \citealt{Vaupre2014}). In this case, the abundances of \dco\ and \hco\ mostly depend on the \ionRate$/n_H$ ratio and not only on \ionRate, and using $R_D$ to get \ionRate\ is valid only when deuteration is $>2\%$: the measured values in B335 are mostly all lower than this threshold at radii $< 500$~au, which would suggest that our approach of using these observations to derive \ionFrac\ is deficient. However, recent models have shown that in warm dense gas (T $\gtrsim$ 24~K and $n_{\rm H} \sim 10^6$ cm$^{-3}$), which is typical of the conditions probed here, the relation between \deutFrac\ and \ionRate\ is more robust. \citet{Shingledecker2016} show that the scatter of \ionRate\ predicted by the Caselli model at low ionization values seems to be related to variations in the ortho-to-para ratio (OPR) of H$_2$. While the OPR is an unobservable parameter for dense gas conditions, and therefore cannot be measured for our source, the work of these latter authors shows that the dependence of the relation between \deutFrac\ and \ionRate\ on OPR is weaker for evolved sources due to the stabilization of the OPR. This effect has been shown to happen sooner when $n_{\rm H}$ and the temperature are increased. This was also pointed out by \citet{Bron2021} who show that \deutFrac\ is a robust estimator of \ionRate\ when the gas is subject to high ionization fractions (\ionFrac $\gtrsim 3 \times 10^{-7}$). As we are probing a dense region of the core with temperatures larger than 20~K and ionization from an embedded source, this supports our use of the Caselli method to measure \ionRate.
        
        The large \ionFrac\ values could be produced by local UV and X-ray radiation produced in outflow shocks, altering the chemistry from the molecular content around the shocks \citep[e.g.,][]{Viti1999, Girart2002, Girart2005}. However, ionization by far-UV (FUV) photons can be ruled out. Indeed, the maximum energy of FUV photons is around 13~\ev, thus below the threshold for ionization of molecular hydrogen (15.44~\ev). 
        The second argument is related to the fact that the extinction cross-section at FUV wavelengths ($\simeq0.1~\mu$m) is $\sigma_{\rm UV}\simeq2\times10^{-21}$~cm$^2$ per hydrogen atom \citep{Draine2003}. Thus, FUV photons are rapidly absorbed in a thin layer of column density equal to $(2\sigma_{\rm UV})^{-1}\simeq3\times10^{20}$~cm$^{-2}$.
        
        The absorption cross-section of X-ray photons, $\sigma_{\rm X}$, in the range 1$-$10~\kev\ is much smaller than at FUV frequencies \citep{Bethell2011}. For example, at 1 keV and 10 keV, $\sigma_{\rm X}\simeq2\times10^{-22}$~cm$^2$ and $\simeq8\times10^{-25}$~cm$^2$
        per H atom, respectively. While at 1~\kev\ the corresponding absorption column density is still small, $(2\sigma_{\rm X})^{-1}\simeq2\times10^{21}$~cm$^{-2}$, at 10~\kev\ it reaches about $6\times10^{23}$~cm$^{-2}$, much larger than the maximum \hydrogen\ column density, ${\bm N_{\rm H_{2}}}$, that we found. The latter was computed by integrating the dust continuum at 110~\Ghz\ along directions identified by the position angle $\vartheta$ (see lower right panel of Fig.~\ref{fig:zeta_vs_radii}) and is representative of the 
        column density in the outflow cavity. We therefore estimated the X-ray ionization rate, \XionRate, using the spectra described by \citet{Rab2017} for a typical and a flaring T Tauri star, assuming a stellar radius of $2R_\odot$. Results for \XionRate\ are shown in the upper left panel of Fig.~\ref{fig:zeta_vs_radii}. It is evident that X-ray ionization cannot explain the values of the ionization rate estimated from observations. A fraction of these X-ray photons end up in the disk, which has much higher densities than the cavity, and so ${\bm N_{\rm H_{2}}}$ can easily reach values greater than $10^{24}$~cm$^{-2}$, as found by \citet{Grosso2020} for a Class 0 protostar. At such high ${\bm N_{\rm H_{2}}}$, \XionRate\ decreases dramatically.

        Once the high-energy radiation field is excluded as the main source of ionization rate estimated from observations, the only alternative left is CRs (see Sect.~\ref{sec:IonRate}). However, the range of derived values is much higher than those expected from Galactic CRs
        \citep[see Appendix C in][for an updated compilation of observational estimates of \ionRate]{Padovani2022}. Following \citet{McKee1989}, \ionFrac\ $=1.3 \times10^{-5} n(\rm H_2)^{-1/2}$, and for the volume densities derived in Sect.~\ref{sec:IonRate}, we should expect \ionFrac\ values of $10^{-7}$ to $10^{-8}$, which are between one and two orders of magnitude lower than the observed values in B335. As we also observed a significant increase in \ionFrac\ toward the center, and especially of \ionRate, the origin of the high values should be of local origin. In particular, we are most likely witnessing the local acceleration of CRs in shocks located in \bttf\ as predicted by theoretical models \citep{Padovani2015, Padovani2016, Padovani2020}. The values of \ionRate\ found in \bttf, between about $10^{-16}$ and $10^{-14}$~s$^{-1}$, are among the highest values reported in the literature toward star-forming regions \citep[e.g.,][]{Maret2007,Ceccarelli2014b,Fuente2016,Fontani2017,Favre2017,Bialy2022}. We stress that this is the first time that \ionRate\ is measured at such small scales in a solar-type protostar and can be attributed to locally accelerated CRs, as well as the first time that a map of \ionRate\ is obtained.
        
        There are two possible origins for the production of local CRs close to a protostar. One is in strong magnetized shocks along the outflow
        \citep{Padovani2015,Padovani2016,Fitz21,Padovani2021}. Indeed, synchrotron radiation, which is the signature of the presence of relativistic electrons, has been detected in some outflows \citep{Carrasco10, Ainsworth14, RodriguezK16, RodriguezK19, Osorio17}. The second possibility is in accretion shocks near the stellar surface \citep{Padovani2016,Gaches2018}. Both mechanisms are expected to generate low-energy CRs ($\lesssim$ 100~\mev - 30~\gev) through the first-order Fermi acceleration mechanism with a rate of up to $\sim10^{-13}$~s$^{-1}$. 

        The ionization rate slope found in \bttf\ for $r<270$~au is close to $-1$ and compatible with a diffusive regime, in agreement with predictions of theoretical models. Surprisingly, at radii larger than $\sim$ 250~au, the slope decreases below $-2$, thus beyond the geometrical dilution regime. We propose two possible explanations. On the one hand, local CRs may have accumulated enough column density to start being thermalized. On the other hand, at radii above about 250~au, both the dust emission and \ionRate\ maps gradually lose their central symmetry (see Figs.~\ref{fig:intensity_maps} and ~\ref{fig:ionRate_maps}, respectively). For example, the continuum map shows two higher density arms toward the northeast and southeast. Therefore, the slope,  calculated by averaging over the position angle distribution $\vartheta$, could be less than $-2$. The loss of symmetry at larger radii in the ionization maps might also be additional evidence as to the importance of non-symmetrical motions during the protostellar collapse \citep{Cabedo2021b}. However, we note that these values are also affected by our limited angular resolution of 1.5~\arcsecond\ ($\sim$ 250~au), implying that only three or four points in the plot are completely independent, and that observations at larger angular resolution are needed in order to confirm the observed trend.

        Previous observations indicate that \bttf\ hosts a powerful and variable jet, which might explain a large production of local CRs \citep{Galfalk2007, Yen2010}. Nevertheless, the exact origin of the local CR source is difficult to pinpoint due to our limited angular resolution. The column density calculated from the dust continuum map at 110~\Ghz\ (see lower left panel of Fig.~\ref{fig:zeta_vs_radii}) is consistent with that expected in the outflow cavity. To calculate the minimum energy that protons must have in order to pass through a given column density without being thermalized, we can make use of the stopping range function (see Fig.~2 in \citealt{Padovani2018}). At ${\bm N_{\rm H_{2}}}\simeq10^{22}$~cm$^{-2}$, protons with energies of about 10~\mev\ are thermalized and lose their ionizing power (see also the upper x-axis of Fig.~\ref{fig:zeta_vs_radii}). 
        As discussed above, if the shock is in the vicinity of the protostellar surface, a column density of the order of $10^{24}-10^{25}$~cm$^{-2}$ can easily be accumulated \citep{Grosso2020}. In this case, the minimum proton energy required to prevent thermalization is 100~\mev\ and 400~\mev\ at $10^{24}$ and $10^{25}$~cm$^{-2}$, respectively. Models of local CR acceleration in low-mass protostars predict that accelerated protons can reach maximum energies of the order of 100~\mev\ and 10~\gev\ if the shock in which they are accelerated is located in the jet or close to the protostellar surface, respectively \citep{Padovani2015,Padovani2016}. Thus, although the position of the shock that accelerates these local CRs in \bttf\ is unknown, models suggest that the maximum energies of the protons are sufficiently high to explain the observed ionization.
        
        Additionally, we note that \bttf\ has been observed to exhibit an organized magnetic field at scales similar to the ones probed here \citep{Maury2018}. This could also create favorable conditions for enhanced CR production. Finally, our results are in agreement with CR ionization models on protostellar cores with an embedded radiation source \citep{Gaches2019}; these latter models also suggest that ionization models based solely on \htp\ chemistry might underpredict \ionFrac\ by one order of magnitude, except in regions where CR ionization is dominant.

    \subsection{Magnetic field coupling and non-ideal MHD effects}
    
        The ionization fraction of the gas in a young accreting protostar is not only important to understand the early chemistry around solar-type stars, feeding the scales where disks and planets will form, but is also the critical parameter that determines $(i)$ 
        the coupling between the magnetic field and the infalling-rotating gas in the envelope, and $(ii)$ the role of diffusive processes, such as ambipolar diffusion, which counteract the outward transport of angular momentum due to magnetic fields (a process known as magnetic braking). The higher the gas ionization, the more efficient the coupling, and the braking.

        The large fraction of ionized gas unveiled by our observations in the inner envelope of \bttf\ should lead to almost perfect coupling of the gas to the local magnetic field lines, generating a drastic braking of rotational motions. We note that these new results lend extra support to other observational evidence of a very efficient magnetic braking at work in \bttf, notable the highly pinched magnetic field lines observed at similar scales by \citet{Maury2018}, and the failure to detect any disk larger than $\sim10$~au in this object \citep{Yen2015b}; although a new kinematic analysis may be motivated in light of the detection of several velocity components in the accreting gas at scales of $\sim500$~au \citep{Cabedo2021b}.

        At the end of the pre-stellar stage, the fraction of ionized gas toward the central part of the core is expected to be very low, as Galactic CRs do not penetrate deeply and there is no local ionization source \citep{Padovani2013,Ceccarelli2014,Silsbee2019}. Hence, the initial stages of protostellar evolution are proceeding under low ionization conditions at typical \ionRate\ $<10^{-16}\,\rm{s}^{-1}$ \citep{Padovani2018,Ivlev2019}, which enhance the importance of non-ideal MHD processes \citep{Padovani2014}, with efficient diffusion of magnetic flux outwards during the very first phases of the collapse, and reduced magnetic braking. If the local ionization processes we observe in \bttf\ are prototypical of solar-type Class 0 protostars, then CRs accelerated in the proximity of the protostar could be responsible for changing the ionization fraction of the gas at disk-forming scales once the protostar is formed, setting quasi-ideal MHD conditions in the inner envelope. The timescale and magnitude of this transition from non-ideal MHD conditions to quasi-ideal MHD conditions may depend on the protostellar properties: more detailed modeling and observations toward other protostars are required to address this question. Moreover, we note that the observed large \ionFrac\ is not in agreement with the values used to calculate the simplified chemical networks in non-ideal MHD models of protostellar formation and evolution \citep{Marchand2016,Zhao2020a}, as gas ionization is usually predicted from the gas density following \citet{McKee1989}. Thus, our observations suggest that careful treatment of ionizing processes in magnetized models may be crucial to properly describe the gas-magnetic field coupling at different scales in embedded protostars.
        
        In this scenario, the properties of protostellar disks could be tightly related to the local acceleration of low-energy CRs, and the development of a highly ionized region around the protostar. Therefore, we would not expect the ionization fraction of the gas present at large scale in the surrounding cloud to be a key factor in setting the disk properties, as proposed for example by \citet{Kuffmeier2020}. The scenario we propose is also in agreement with recent ALMA observations of the Class II disks in Orion that do not find supporting evidence of local cloud properties affecting the disk properties \citep{vanTerwisga2022}.

\section{Conclusions and summary} \label{sec:Conclusions}

    This work provides new insights into the physico-chemical conditions of the gas in the young Class 0 protostar \bttf. For the first time, we derived a map of the gas \ionFrac\ and of \ionRate\ at envelope scales $<1000$~au in a Class 0 protostar. Our results highlight the interplay between physical processes responsible for gas ionization at disk-forming scales, and its consequences for magnetized models of solar-type star formation. Here, we summarize the main results of our analysis:
        
        \begin{itemize}
        
            \item We used ALMA data to create molecular line emission maps and used these to characterize the physico-chemical properties of the dense gas in B335. We used the chemical recipe from \citet{Caselli1998} to derive \ionFrac\  based on the deuteration fraction of HCO+ and of the \deplFact. This recipe was recently shown to be valid for dense molecular gas, $\sim 10^6$ cm$^{-3}$,  with a high ionization fraction ($\gtrsim 3 \times 10^{-7}$, \citealt{Bron2021}). It allows us to characterize the ionization of the gas at envelope radii $\lesssim$ 1000~au: we find large fractions of ionized gas, \ionFrac, between 1 and $8\times10^{-6}$. These values are remarkably higher than the ones usually measured at core scales.  
            
            \item We produced for the first time a map of \ionRate, in the close environment of a young embedded protostar. Our analysis reveals very high values of \ionRate\  between $10^{-16}$ and $10^{-14}$~s$^{-1}$, increasing toward the central protostellar embryo at small envelope radii. This suggests that local acceleration of CRs, and not the penetration of interstellar CRs, may be responsible for the gas ionization in the inner envelope, potentially down to disk forming scales.
            
            \item The large fraction of ionized gas we find suggests an efficient coupling between the magnetic field and the gas in the inner envelope of \bttf. This interpretation is also supported by the observations of highly organized magnetic field lines, and no detection of a large rotationally supported disk in \bttf.
            
        \end{itemize}
        
        If our findings are found to be prototypical of the low-mass star formation process, they might imply that the collapse at scales $<1000$~au transitions from non-ideal to a quasi-ideal MHD once the central protostar starts ionizing its surrounding gas, and very efficient magnetic braking of the rotating-infalling protostellar gas might then take place. Protostellar disk properties may therefore be determined by local processes setting the magnetic field coupling, and not only by the amount of angular momentum available at large envelope scales and by the magnetic field strength in protostellar cores. We stress that the gas ionization we find in \bttf\ significantly stands out from the typical values routinely used in state-of-the-art models of protostellar formation and evolution. Our observations suggest that a careful treatment of ionizing processes in these magnetized models may be crucial to properly describe the processes responsible for disk formation and early evolution. We also note that more observations of \bttf\ at higher spatial resolution, and of other Class 0 protostars, are crucial to confirm our results.

\section*{Acknowledgments}

This project has received funding from the European Research Council (ERC) under the European Union Horizon 2020 research and innovation programme (MagneticYSOs project, grant agreement N$\degr$ 679937). This work was also partially supported by the program Unidad de Excelencia María de Maeztu CEX2020-001058-M. JMG also acknowledges support by the grant PID2020-117710GB-I00 (MCI-AEI-FEDER, UE).

Additionally, this paper makes use of the following ALMA data: \YenProj. ALMA is a partnership of ESO (representing its member states), NSF (USA) and NINS (Japan), together with NRC (Canada), MOST and ASIAA (Taiwan), and KASI (Republic of Korea), in cooperation with the Republic of Chile. The Joint ALMA Observatory is operated by ESO, AUI/NRAO and NAOJ


\bibliographystyle{aa}
\bibliography{mybiblio.bib}

\appendix

\section{Technical details of the ALMA observations} \label{sec:ap_ObsDetails}

    Table~\ref{table:technicalInfo} presents the technical details of the ALMA observations presented in this work, including the ALMA configuration, the date and time range, the central frequency, the obtained spectral resolution, and the different calibrators used.

    \begin{table*}[!ht]
    \centering
    \caption{Technical details of the ALMA observations.}
    \begin{tabular}{c c c c c c c c}
        \toprule\toprule
        Config. & Date & Timerange & 
        \multicolumn{1}{c}{Center freq}. & Spec. Res. & Flux cal. & Phase cal. & Bandpass cal.\\
        & (mm/dd/yy) & (hh:min:s.ms; UTC) & (GHz) & (km s$^{-1}$) & &  &   \\
        \midrule
        C40-1 & 19/03/2017 & 10:40:52.8 - 10:53:42.8 & 231.0 & 0.09 & J1751+0939 & J1851+0035 & J1751+0939 \\
        C40-2 & 02/07/2017 & 15:30:38.9 - 15:49:59.2 & 110.0 & 0.08 & J1751+0939 & J1938+0448 & J1751+0939 \\
        C40-3 & 12/06/2016 & 18:39:47.5 - 19:24:24.3 & 87.7 & 0.10 & J1751+0939 & J1938+0448 & J1751+0939 \\
        C40-4 & 21/11/2016 & 22:41:22.0 - 23:28:19.8 & 231.0 & 0.09 & J2148+0657 & J1851+0035 & J2148+0657 \\
        C40-5 & 10/22/2016 - 10/23/2016 & 23:34:05.4 - 00:02:51.3 & 110.0 & 0.08 & J2148+0657 & J1938+0448 & J2025+3343 \\
        C40-6 & 10/10/2016 & 20:50:38.0 - 21:39:40.2 & 87.7 & 0.10 & J1751+0939 & J1938+0448 & J1751+0939 \\
         & 10/10/2016 & 22:43:07.4 - 23:24:00.4 & 87.7 & 0.10 & J1751+0939 & J1938+0448 & J1751+0939 \\
         & 10/11/2016 & 21:14:59.4 - 22:03:55.2 & 87.7 & 0.10 & J1751+0939 & J1938+0448 & J1751+0939 \\
        \bottomrule
    \end{tabular}
    \label{table:technicalInfo}
\end{table*}

\section{Line profile analysis}

    \subsection{Spectral maps} \label{sec:ap_spectralMaps}

\begin{figure*}[!ht]
            \centering
            \includegraphics[width=0.99\textwidth]{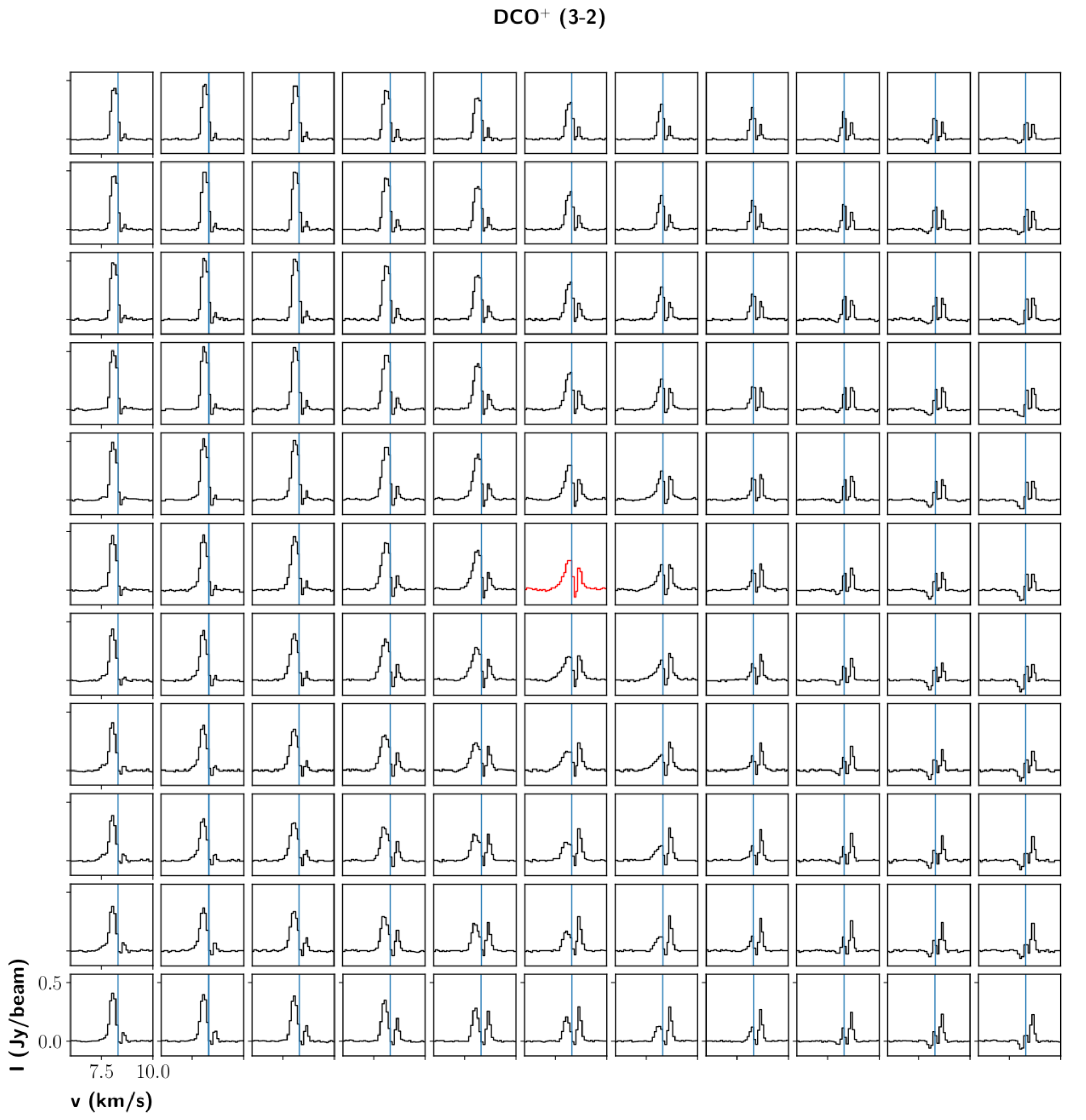}
            \caption{Spectral map of the \dco\ (J=3-2) emission in the inner 900~au, centered on the dust continuum emission peak. The whole map is 5.5\arcsec\ $\times$ 5.5\arcsec\ and each pixel corresponds to 0.5\arcsec ($\sim$ 82~au). The red spectrum refers to the peak of the continuum emission and the blue line indicates the systemic velocity (8.3~\kms).}
            \label{fig:dco_spectralMap}
        \end{figure*}

        The \dco\ (J=3-2) and \htco\ (J=3-2) spectral maps are presented in Figs.~\ref{fig:dco_spectralMap} and \ref{fig:h13co_32_spectralMap}. Both maps show emission in a region of 5.5\arcsec\ $\times$ 5.5\arcsec\ (900~au) around the center of the source, with each spectrum corresponding to a pixel of 0.5\arcsec\ (82~au). The \htco\ (J=1-0) shows only a region 4\arcsec\ $\times$ 4\arcsec\ (685~au) and each spectrum corresponds to 0.25\arcsec\ (41~au). 
        
        The three tracers present the same pattern observed in \coseven\ (J=1-0), presented in \citet{Cabedo2021b}. The line profiles are double-peaked with one of the two velocity components progressively disappearing at different offsets of the source, that is, the blueshifted component is dominant in the eastern region of the source while the redshifted component is present only in the western region. For the three transitions, the blueshifted component is always more intense than the redshifted one, which produces the blue asymmetric morphology observed in the integrated intensity images. 
        
        As none of the tracers present a morphology typical of the outflow cavity, we assume that they are not tracing outflowing gas and that the two velocity components are not caused by this effect. This is confirmed with the spectral maps where no signatures coming from outflowing gas ---such as very broad lines or high velocity wings--- can be observed in the line profiles. This is generally true, except for a small region on the southeast region of the \htco\ (J=3-2) map, where an additional third velocity component, or "shoulder", can be observed. This additional component appears to be very broad, and at velocities of $\sim$7.6-–7.7~\kms. We do attribute this signature to outflow contamination, which is observed neither in \dco\ (J=3-2) nor in \htco\ (J=1-0). We consider that while the outflow might have some effect in the line profiles, this is negligible and is not the main cause of the double-peaked line profiles. The effect of large-scale filtering cannot be ruled out. This is particularly visible in the \dco\ (J=3-2) and \htco\ (J=3-2) where negative emission is observed in the center and west regions of the source. While these two effects can be a source of uncertainty in our analysis, we conclude that they alone cannot produce the observed line profiles and the velocity pattern through the source, and proceed with the line modeling with the assumption of two independent velocity components.
    
        \begin{figure*}
            \centering
            \includegraphics[width=0.99\textwidth]{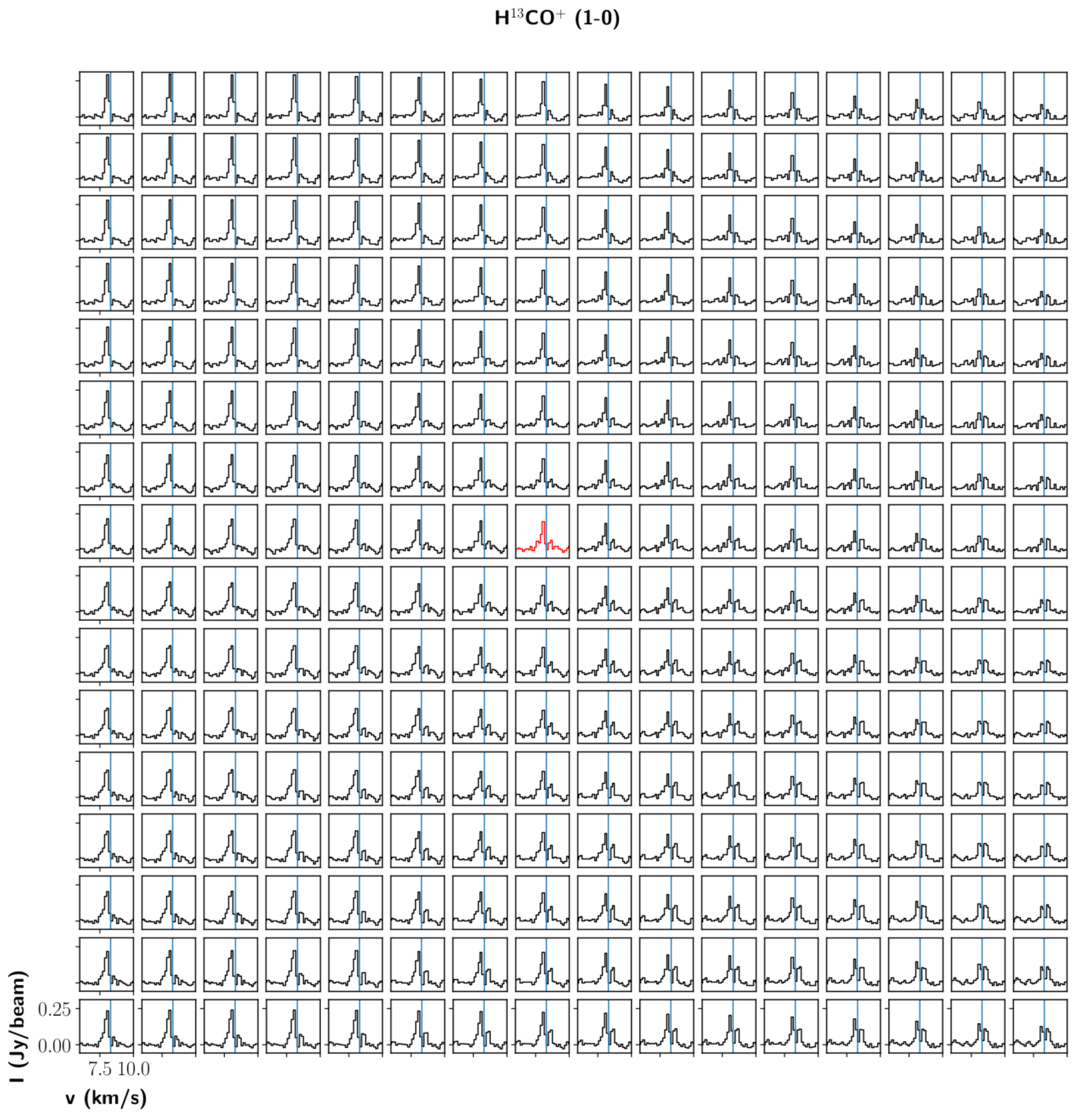}
            \caption{Spectral map of the \htco\ (J=1-0) emission in the inner 658~au centered on the dust continuum emission peak. The whole map is 4\arcsec\ $\times$ 4\arcsec\ and each pixel corresponds to 0.25\arcsec\ ($\sim$ 41~au). The red spectrum refers to the peak of the continuum emission, and the blue line indicates the systemic velocity (8.3~\kms).}
            \label{fig:h13co_10_spectralMap}
        \end{figure*}
        
        \begin{figure*}
            \centering
            \includegraphics[width=0.99\textwidth]{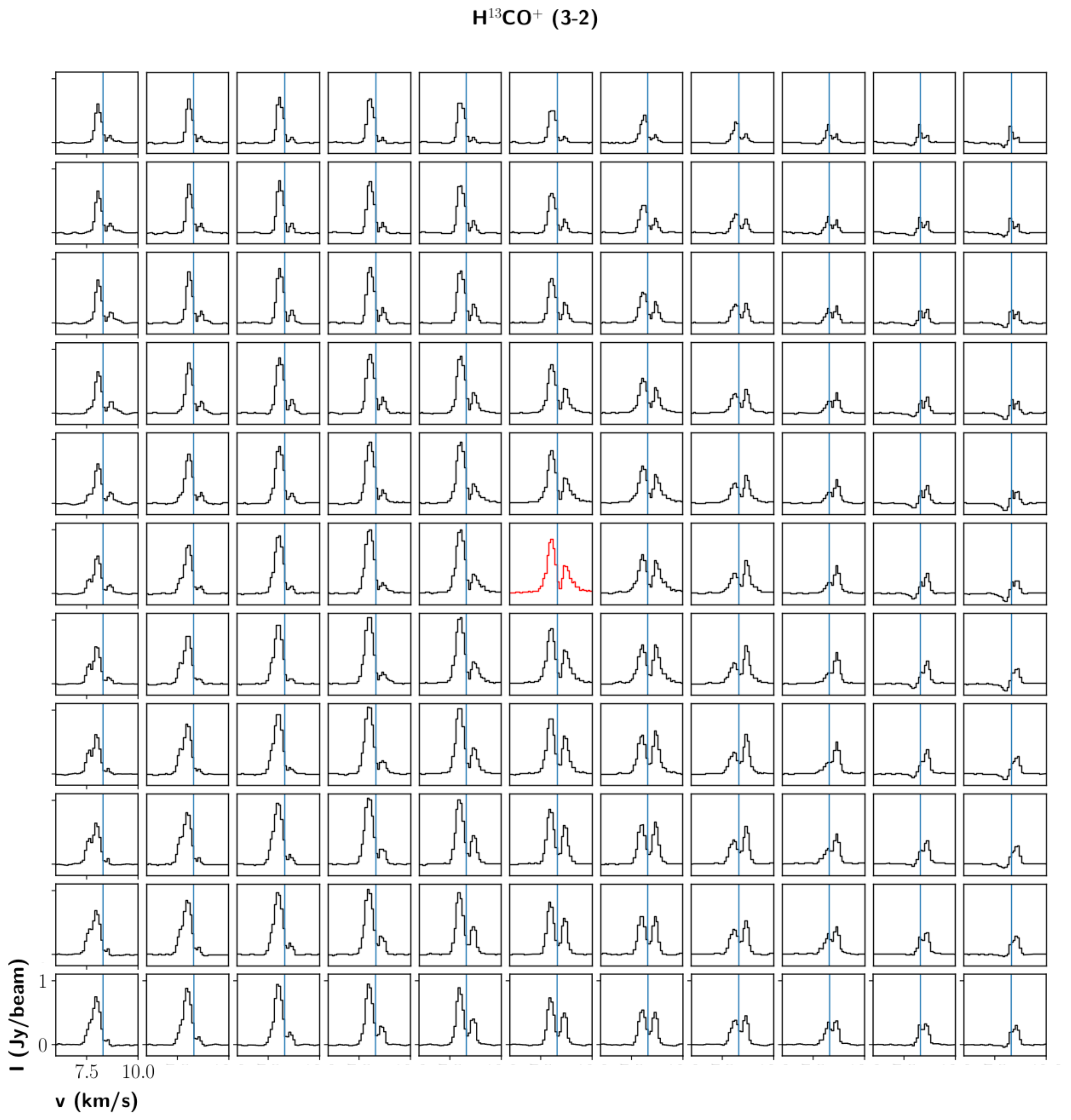}
            \caption{Spectral map of the \htco\ (J=3-2) emission in the inner 900~au centered on the dust continuum emission peak. The whole map is 5.5\arcsec\ $\times$ 5.5\arcsec\ and each pixel corresponds to 0.5\arcsec ($\sim$ 82~au). The red spectrum refers to the peak of the continuum emission and the blue line indicates the systemic velocity (8.3~\kms)}
            \label{fig:h13co_32_spectralMap}
        \end{figure*}
    
    \subsection{Line profile modeling} \label{sec:ap_LineMod}
    
\begin{figure*}
            \centering
            \includegraphics[width=0.99\textwidth]{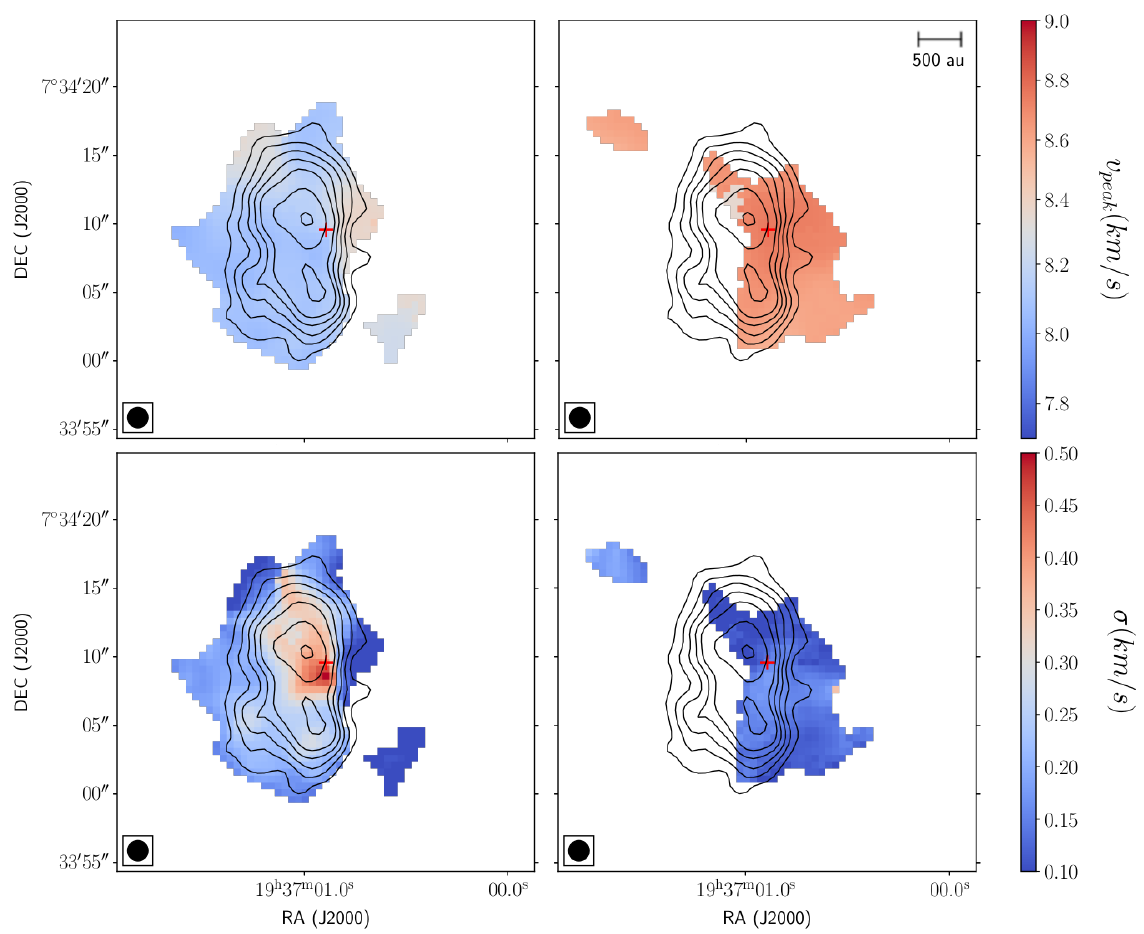}
            \caption{\dco\ (J=3-2) maps obtained from modeling the line profiles with two velocity components. Overlaid contours show the integrated intensity at -3, 3, 5, 10, and 20 $\sigma$. The top map shows values for the peak velocity and bottom show values for the line width (given as $\sigma$). Left and right show the blue- and redshifted component, respectively.}
            \label{fig:dco_lineMod}
        \end{figure*}

\begin{figure*}
            \centering
            \includegraphics[width=0.99\textwidth]{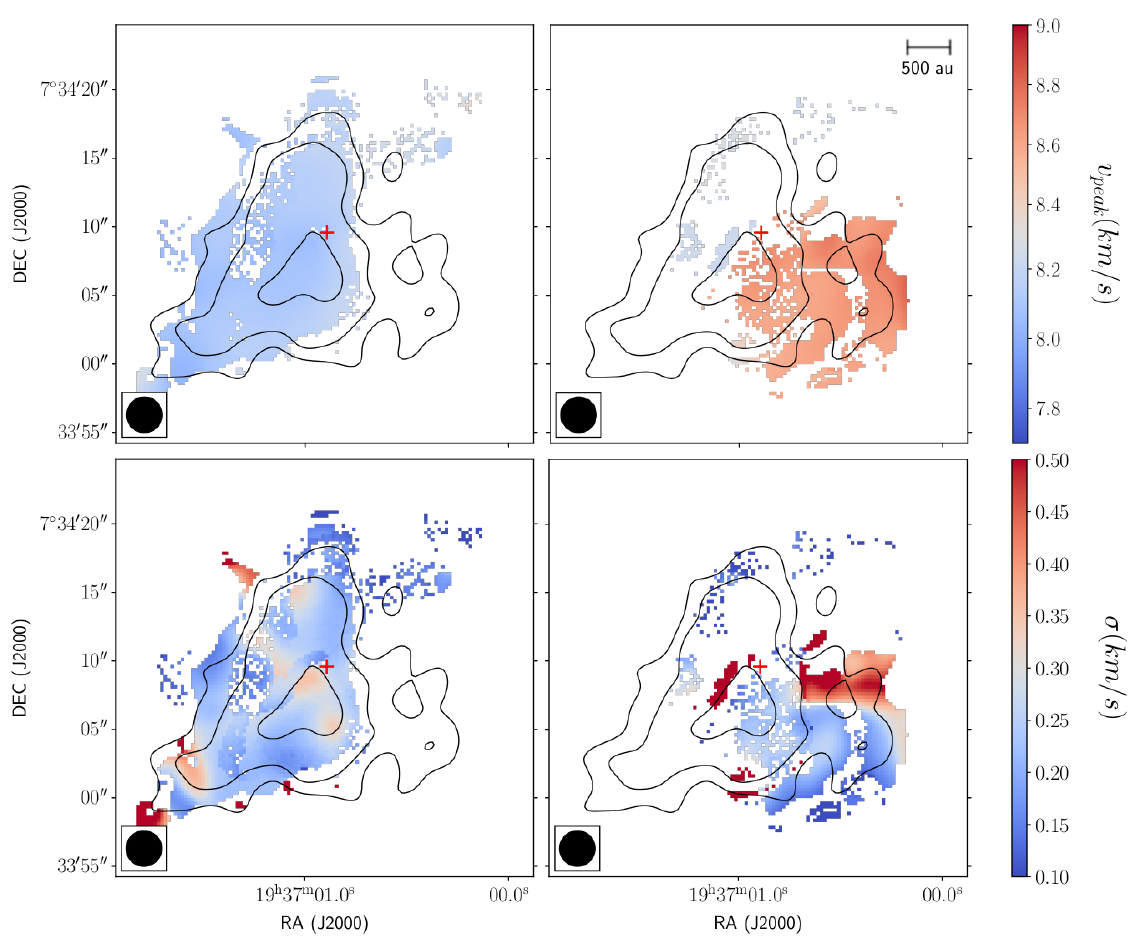}
            \caption{\htco\ (J=1-0) maps obtained from modeling the line profiles with two velocity components. Overlaid contours show the integrated intensity at -3, 3, 5, and 10 $\sigma$. The top map shows values for the peak velocity and bottom show values for the line width (given as $\sigma$). Left and right show the blue- and redshifted component, respectively.}
            \label{fig:h13co_10_lineMod}
        \end{figure*}

\begin{figure*}
            \centering
            \includegraphics[width=0.99\textwidth]{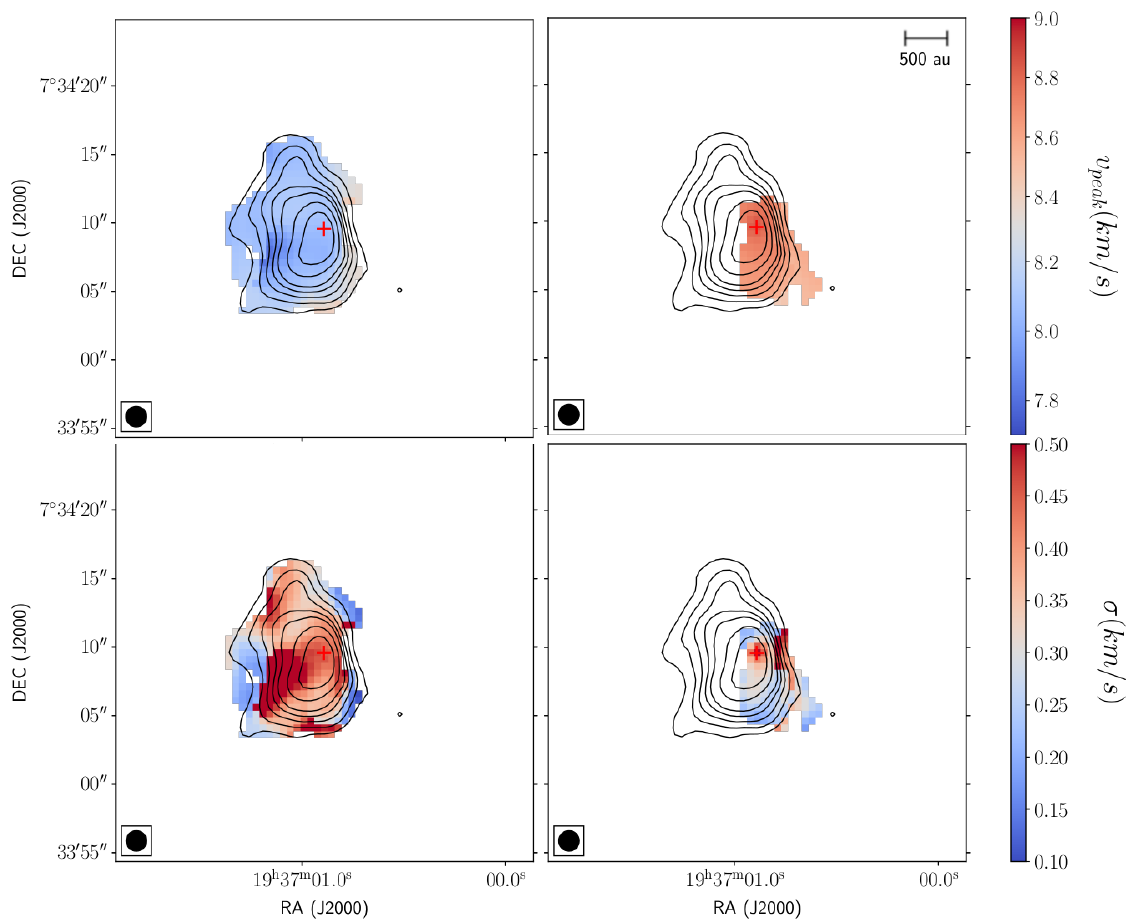}
            \caption{\htco\ (J=3-2) maps obtained from modeling the line profiles with two velocity components. Overlaid contours show the integrated intensity at -3, 3, 5, 10, and 20 $\sigma$. The top map shows values for the peak velocity and bottom show values for the line width (given as $\sigma$). Left and right show the blue- and redshifted component, respectively.}
            \label{fig:h13co_32_lineMod}
        \end{figure*}

        In this case, as none of the transitions present a resolved hyperfine structure, the opacity of the line could not be obtained by modeling the line profile and only the peak velocity and velocity dispersion maps (as $\sigma$) are presented. The resulting maps for \dco\ (J=3-2), \htco\ (J=1-0), and \htco\ (J=3-2) are shown in Figs.~\ref{fig:dco_lineMod}, \ref{fig:h13co_10_lineMod}, and \ref{fig:h13co_32_lineMod}. For the three tracers, the two separated velocity components occupy different regions of the source, indicating that they are probing gas with different motions. The morphology of the two components is similar for the three tracers, but it is slightly different from the fitting obtained with the \coseven\ (Fig. 5 in \citealt{Cabedo2021b}), where the redshifted component is almost always present in the \coseven\ spectra.
        
        Peak velocity value (top panels of Figs.~\ref{fig:dco_lineMod}, \ref{fig:h13co_10_lineMod} and \ref{fig:h13co_32_lineMod}) ranges are similar for all molecules (from $\sim$8 to 9~\kms), although the blueshifted component goes to slightly lower velocities for \htco\ (J=3-2), being the lowest velocities $\sim$7.8~\kms. This decrease occurs toward the southeastern region of the core, at $\sim$5\arcsec\ from the center. We attribute this to the presence of the third velocity component produced by outflow contamination observed in Fig.~\ref{fig:h13co_32_spectralMap}, which produces a bad model of the line profile. No gradient indicating rotation is observed in the velocity maps for any tracer.
        
        Velocity dispersion (bottom panels of Figs.~\ref{fig:dco_lineMod}, \ref{fig:h13co_10_lineMod} and \ref{fig:h13co_32_lineMod}) ranges are also similar for all tracers (from 0.1 to 0.5~\kms). However, there are some main differences: velocity dispersion is generally larger for \htco\ (J=3-2) with average values of $\sim$0.4~\kms, whereas for \dco\ (J=3-2) and \htco\ (J=1-0) mean values are around $\sim$ 0.25~\kms. Very large values of the velocity dispersion ($>$0.5 \kms) are observed in the same southeastern region of the blueshifted component of \htco\ (J=3-2) which is a consequence of the poor fitting due to the third velocity component. Values for the velocity dispersion are generally larger for the blueshifted than for the redshifted velocity component, indicating the two components trace gas with different kinematics, i.e., different infall velocities are affected differently by turbulence. For all tracers and both velocity components, the velocity dispersion increases toward the center, as the temperature rises. This increase is smaller for the \htco\ (J=1-0) emission (up to $\sim$0.35~\kms) than for the \dco\ (J=3-2) and \htco\ (J=3-2) (up to $\sim$0.5~\kms) indicating that \htco\ (J=1-0) traces gas at colder temperatures and further to the center of the source.
        
        In general, there is good agreement between the values obtained for \dco\ (J=3-2) and \htco\ (J=3-2), both for peak velocity and velocity components. Considering that we are tracing similar scales, this confirms that both tracers are tracing similar gas. However, \coseven\ presents some differences with respect to \htco\ (J=1-0) which indicates that they are tracing slightly different gas, \coseven\ tracing warmer and denser gas closer to the protostar.
        
\section{Estimation of the opacity effects}\label{ssec:Opacity}
    
\begin{figure*}
            \centering
            \includegraphics[width=\textwidth]{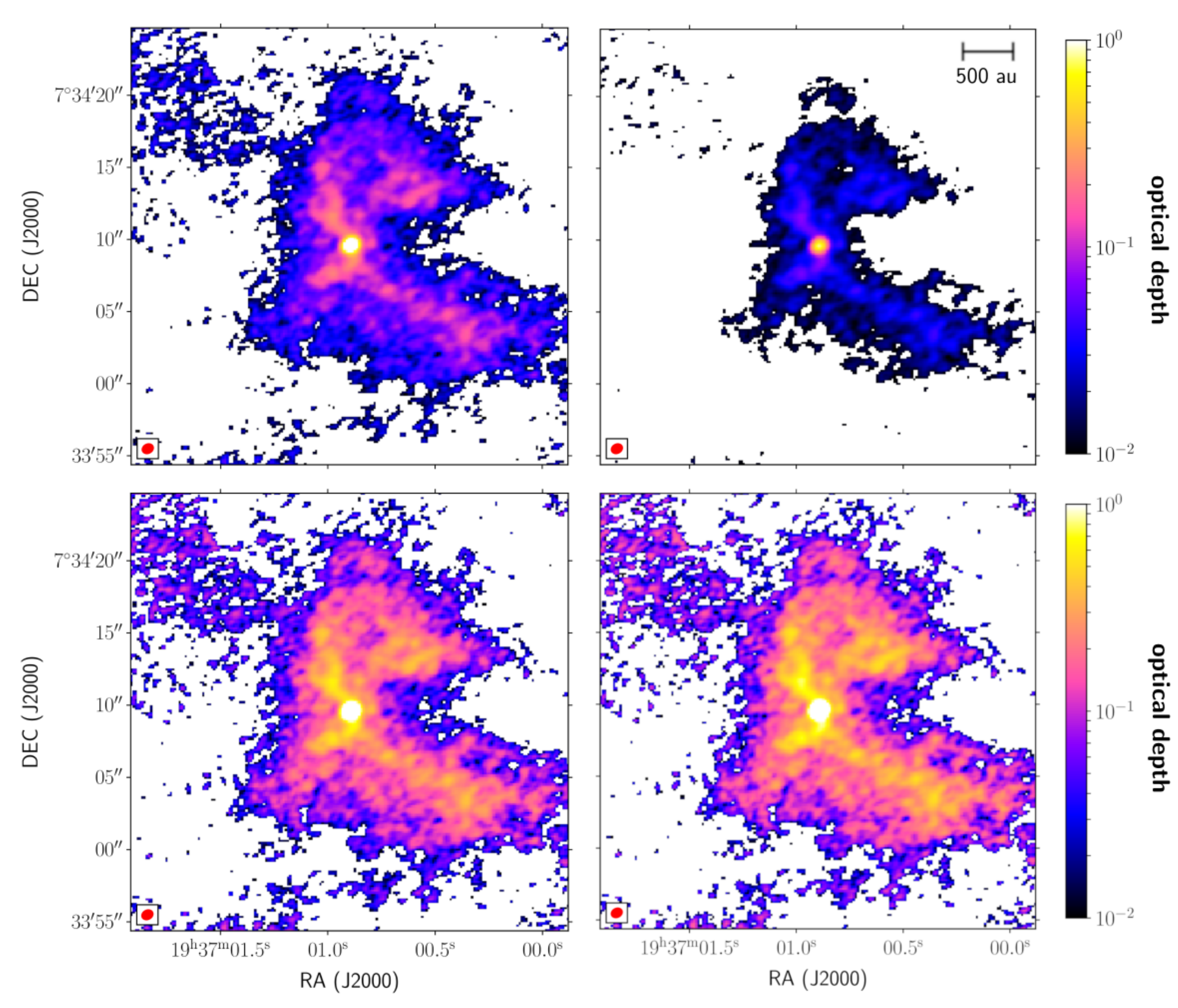}
            \caption{Expected optical depth maps for the four molecular tracers (see text in Section \ref{ssec:Opacity} for details). \textit{Top left:} \coseven\ (J=1-0). \textit{Top right:} \dco\ (J=3-2). \textit{Bottom left:} \htco\ (J=1-0). \textit{Bottom right:} \htco\ (J=3-2).}
            \label{fig:opacities_110GhzCont}
        \end{figure*}

        As the line profile modeling does not allow us to obtain the line opacities, we estimated the optical depth using the dust continuum emission and the molecular abundances in order to determine the impact on the parameters derived from the line modeling. We first compute the \hydrogen\ column density map from the dust continuum emission map at 110~\Ghz\ (shown in Fig.~\ref{fig:intensity_maps}), assuming its emission is optically thin (e.g., tracing the full column density) and applying the following standard abundances: [\coseven] = 5$\times$10$^{-8}$ \citep{Thomas2008}, [\dco] = 10$^{-11}$ and [\htco] = 10$^{-10}$ \citep{Jorgensen2004}. Figure~\ref{fig:opacities_110GhzCont} shows the expected optical depth maps for the four different molecular lines we obtain with this method. Values of the optical depth are generally low ($<$ 1), although they increase to larger values for all tracers in the central beam. We note that, as we are considering a constant temperature for the derivation of the \hydrogen\ column density, which is likely higher at the center, the \hydrogen\ column density values in the central compact source are probably smaller, and we are overestimating the optical depth in that region. It can be seen that both transitions of \htco\ present larger values of optical depth than both \coseven\ (J=1-0) and \dco\ (J=3-2). We acknowledge that these large values can potentially affect our deuteration measurements toward the central beam, but stress that the small optical depth values found in \dco\ confirm that the two velocity components are not mainly produced by self-absorption effects. Outside of that central beam, the expected line opacities are generally smaller than 1, so we found no need for applying an opacity correction to our data.
    
\end{document}